\documentclass{aa}
\usepackage{epsfig}
\nonstopmode
\begin{document}

\title{Line emission from optically thick relativistic accretion tori} 

\authorrunning{Fuerst and Wu}
\titlerunning{Line Emission from Optically Thick Tori}

   \author{ Steven V.~Fuerst\inst{1,2} \and Kinwah Wu\inst{2}
}

   \offprints{S.~V.~Fuerst}

   \institute{Kavli Institute for Particle Astrophysics and Cosmology, 
              Stanford University, Stanford, CA 94305, USA \\
              \email{sfuerst@stanford.edu}       
   	\and
   	      Mullard Space Science Laboratory, University College London,
              Holmbury St Mary, Surrey RH5 6NT, UK \\
              \email{kw@mssl.ucl.ac.uk}       
             }

\date{Received:    }

\abstract{ We calculate line emission from relativistic accretion tori around Kerr black holes  
   and investigate how the line profiles depend on the viewing inclination,  
   spin of the central black hole, parameters describing the shape of the tori, 
   and spatial distribution of line emissivity on the torus surface.  
We also compare the lines with those from thin accretion disks.  
Our calculations show that lines from tori and lines from thin disks 
   share several common features. 
In particular, at low and moderate viewing inclination angles  
   they both have asymmetric double-peaked profiles 
   with a tall, sharp blue peak and a shorter red peak which has an extensive red wing. 
At high viewing inclination angles 
    they both have very broad, asymmetric lines which can be roughly considered as single-peaked.     
Torus and disk lines may show very different red and blue line wings, 
   but the differences are due to the models for relativistic tori and disks having differing inner boundary radii.  
Self-eclipse and lensing play some role in shaping the torus lines, 
   but they are effective only at high inclination angles. 
If inner and outer radii of an accretion torus are the same as those of an accretion disk, 
   their line profiles show substantial differences only when inclination angles are close to $90^\circ$, 
   and those differences manifest mostly at the central regions of the lines instead of the wings. 
 }

   \maketitle
%

\newcommand{\etal}{{et al.}\ }
\newcommand{\eg}{e.g.\ }
\newcommand{\ie}{i.e.\ }

\def\Eqn#1{(\ref{#1})}

\newcommand{\RG}{R_{\rm g}}

\section{Introduction}

The keV X-ray spectrum of active galactic nuclei (AGN) and black hole X-ray binaries 
   often contains a power-law component and a reflection component, 
   which have lines and edges. 
The power-law continuum arises 
   when soft thermal photons from the accretion disk are Compton up-scattered 
   by hot electrons in a disk corona (Sunyaev \& Titarchuk 1980). 
The reflection component is formed when the hard photons of the power-law component 
   presumably from the hot corona are reflected from the cooler surface of the accretion disk 
   (George \& Fabian 1991; Magdziarz \& Zdziarski 1995).  
The lines of the reflection emission are intrinsically narrow because of their narrow thermal widths.  
They can, however, be broadened by scattering, kinematic energy shifts and gravitational redshift.   
   
Emission lines from a geometrically thin accretion disks
   are expected to have a symmetric double-peaked profile (Smak 1969; Huang 1972). 
The peaks correspond to emission from the two halves of the disk 
   which have opposite line-of-sight velocities.    
Double-peaked lines have been observed in the optical spectra of black-hole X-ray binaries 
   (e.g.\ A0620$-$00, Johnston, Kulkarni \& Oke 1989; GX 339$-$4, Wu \etal 2001),  
   and cataclysmic variables (e.g.\ Z Cha, Horne \& Marsh 1986).  
Some AGN were found to show double-peaked keV X-ray  lines, 
   which have very broad, asymmetric profiles        
   (e.g.\  the Fe K$\alpha$ line in MCG-6-30-15, Tanaka et al.\ 1995).
The blue peak is tall, narrow and sharp,  
   but the red peak is short, with an extensive low-energy (red) wing. 
Moreover, the line centroid appears to be shifted to a lower energy (redshifted). 
These lines are usually interpreted as emission from material 
   circulating with high speeds in the inner accretion disk
   close to the event horizon of the central black hole.  
The unequal peak heights are caused by relativistic boosting 
   --- the intensity of the emission from the approaching flow is enhanced 
   and that of the emission from the receding flow is suppressed. 
The redshift of the line centroid energy and the broadening of the red line wing   
    are consequences of time dilation, 
    a combined effect of transverse Doppler motion, 
    when the flows are in relativistic speeds,  
    and gravitational redshift, 
    as the line photons are required to climb out of the deep gravitational well of the central black hole.  
(For reviews see Fabian \etal (1996, 2000) and references in there.)        
By modelling the observed profiles of the lines in the X-ray spectrum, 
  we may deduce the viewing inclination of the accretion disk, 
  the spatial distribution line emission on the accretion disk 
  and the spin parameter of the accreting black hole. 
  
Calculations of emission lines from geometrically thin relativistic accretion disks/annuli 
   around black holes have been carried out by many workers 
  (e.g.\ Cunningham 1975, 1976;  Gerbal \& Pelat 1981; Fabian et al. 1989; Stella 1990; 
    Kojima 1991; Laor 1991; Bao 1992; Viergutz 1993; Bao, Hadrava \& Ostgaard 1994b; 
   Bromley, Chen \& Miller 1997;  Dabrowski et al.\ 1997; Fanton \etal 1997; 
   Hollywood \& Melia 1997; Cadez, Fanton \& Calvani 1998; Pariev \& Bromley 1998; Reynolds \etal 1999; 
   Fuerst \& Wu 2004; Fuerst 2005; Wu et al.\ 2006).  
The three common methods 
   to calculate the profiles of emission lines 
   from accretion disks 
   are the transfer function method (Cunningham 1975, 1976; Dov\v{c}iak \etal 2004a, 2004b; Czerny \etal 2004), 
   elliptic function method (Bao \etal 1994; Rauch \& Blandford 1994; Agol 1997; Fanton \etal 1997)  
   and direct geodesic integration method (Karas \etal 1992; Reynolds \etal 1999).  
Here we briefly assess their applicability for calculations of profiles of 
   emission lines from thick relativistic accretion tori in turn. 

The transfer function method tabulates a function 
  that specifies the mapping of the surface emitting elements on the accretion disk 
  to the corresponding elements on the sky plane viewed by a distant observer.  
The mapping is not always one to one because of the presence of multiple image orders.  
In practice, only one function per image order is often considered, 
   and some form of hidden surface removal is applied in the calculations.  
The method can, in principle, be used to calculate line profiles for relativistic tori. 
However, one needs to search for a suitable transfer function, 
   which can be complicated, to map the torus surfaces onto the sky plane,   
   making the method difficult for real applications.  
The elliptical function method 
  evaluates the analytic solution of the photon geodesics 
  that links  the emission surface of the accretion disk to the observer.  
It works well for infinitely thin disks, as the disk boundary conditions are simple.   
When self-occultation occurs for images of differing order, 
   a painter's algorithm is used to determine what is visible (see Beckwith \& Done 2004).  
The method is not always applicable for accretion tori. 
For 3D objects the whole path from the emitter to the observer must be checked 
   for possible intersections with another emission surface. 
To do so, one may need to consider a direct integration of the photon geodesics,  
   which is the essence of the third method. 
The direct integration of the geodesic is generally a brute force approach.  
It is not restricted by the specified conditions required 
    by the transfer function and the elliptical function methods. 
The method works well for 3D objects with most boundary conditions,  
   provided that the step size in the integration is small enough 
   (see \eg Fuerst \& Wu 2004).  
In practical calculations, a ray-tracing algorithm is often used.
(See Falcke \etal 2000 for the use of this technique to investigate the potentially VLBI-visible shadow of the black hole in the galactic center.)

The thin disk assumption, which is often used in relativistic disk line calculations, 
   breaks down if the accretion rate is very high. 
Near the Eddington accretion limit, 
   radiation pressure dominates gas pressure in the accretion flows, 
   and gravity is balanced by radiation pressure forces.  
The inner accretion disk may inflate into a thick accretion torus ( see Abramowicz \etal 1978).  
It has been argued that hot coronae are developed above the surfaces of accretion disks. 
For the same reasons,  a hot coronal layer would be present, 
  enveloping the accretion torus.  
The hot coronal layer also gives rise to hard Comptonised photons. 
Provided that the temperature and ionization parameters are low enough,  
   incidence of the Comptonised photons on to the torus could produce fluorescent lines.  
Alternatively, if a hot temperature inversion layer develops on the surface of the accretion torus 
   and the gas in the layer is partially ionised, 
   line emission will emerge.  
   
In this study we calculate the profiles of emission lines arising from a thin, hot surface layer 
   on optically thick relativistic accretion tori 
   and investigate how the torus geometry, combined with relativistic effects,  
   shapes the line profiles.  
We use the direct geodesic integration method (the third method) 
  and employ a ray-tracing algorithm in the calculations. 
We organise the paper as follows. 
In \S2 we review models for accretion tori;  
   in \S3 we construct models for the emission line calculations based on parametrising the angular velocity distribution. 
In \S4 we present the results and compare them
   with those of the case of thin Keplerian accretion disks.

\section{Accretion Torus Model}

The thin disk solution for accretion breaks down 
   when the radiation pressure in the disk dominates the gas pressure 
   and the  radiative pressure force balances the gravitational force.  
   
This happens at very high accretion rate when $\dot{M}$ approaches the Eddington limit.  
For spherical accretion, the corresponding Eddington luminosity is given by
\begin{equation}
   L_{\rm Edd}=\frac{4\pi G M m_{\rm p} c}{\sigma_{\rm T}} \  , 
\label{Ledd}\label{mp}\label{sigmaT}
\end{equation}
  where $G$ is the gravitational constant, $c$ is the speed of light, 
  $m_{\rm p}$ is the proton mass, ${\sigma_{\rm T}}$ is the Thomson cross section, 
  $M$ is the mass of the accreting object. 
Defining the accretion rate that produces this luminosity as $\dot{M}_{\rm crit}$, 
   the scale height of the disk, $H$, can be expressed as 
\begin{equation}
H\simeq\frac{3 R_\star}{4\eta}
  \frac{\dot{M}}{\dot{M}_{\rm crit}}\left[1-\sqrt{\frac{R_\star}{R}}\ \right] \ ,
\label{Mcrit}\label{etaeff}
\end{equation}
   where $\eta$ is the accretion efficiency parameter 
   and $R_\star$ is the radius where angular momentum stops being transported outwards, 
   \ie  the effective inner edge of the disk.  
The disk scale height therefore increases with the accretion rate.     
For $R\gg R_\star$, we have the disk scale height $H\ll R$. 
However, when the accretion rate approaches the Eddington limit, 
   the disk scale height is not negligible in comparison with $R$ 
   in the inner disk, where $R\approx R_\star$.  
Thus, the accretion disk is no longer thin in its inner part.  
In AGN with a black hole accreting at a rate close to the Eddington limit, 
  the inner accretion disk would be geometrically thick, resembling a torus. 
(See \eg Frank, King \& Raine (1985) for a discussion of thick accretion disks.)

In the models of axisymmetric thin disks, 
   the radial component $R$ of cylindrical coordinates 
   is sufficient to describe the field quantities and variables.  
For thick disks, 
   the field quantities and variables are expressed in terms of two components, $R$ and $z$, 
   in cylindrical coordinates, 
   as the vertically-integrated quantities, that are commonly used in the thin disk calculations, 
   are no longer physically meaningful.  
Furthermore, because the $H\ll R$ condition is violated, 
   the Sunyaev-Shakura $\alpha$ viscosity prescription is not applicable. 
Without the $\alpha$ viscosity prescription, 
   the system of equations for the disk hydrodynamics is not longer closed. 
A proper treatment with explicit consideration of the viscosity is complicated,  
  because non-local interaction, 
  \eg magneto-rotational instability can lead to non-local transport of angular momentum in the flow 
  (Hawley, Gammie \& Balbus 1995, Balbus \& Hawley 1998).
   
As the purpose this work is to demonstrate geometrical effects 
  on line profiles in relativistic disks, 
  we may ignore such complications in the treatment of angular momentum redistribution.
There are two methods explored in the literature that can be used to describe the kinematics of these thick disks / tori.  One such method is to parametrise the angular momentum as a function of position within the torus. (Abramowicz \etal 1978, 2004).  Another is to parametrise the angular velocity (Fuerst \& Wu 2004).  The parametrisation is justified if the energy dissipated in the torus can flow in any direction before it is radiated from the torus surface.  
We note that in this case the radiative flux cannot be neatly separated into radial and vertical components as in the thin $\alpha$-disk model.

In this paper we investigate the properties of the tori in Kerr space-time parametrised by their angular velocity.  If this is done, then one may obtain the equations
\begin{eqnarray}
  \frac{d r_{\rm surf}}{d \lambda}
    &=& \frac{\beta_1}{\sqrt{\beta_2^2+\Delta\beta_1^2}}\ ,   \nonumber  \\
  \frac{d \theta_{\rm surf}}{d \lambda}
    &=& \frac{-\beta_2}{\sqrt{\beta_2^2+\Delta\beta_1^2}}\ ,  
\label{isobaric} 
\end{eqnarray}
where
\begin{eqnarray}
   \beta_1&=&\frac{\Sigma-2r^2}{\Sigma^2}
   \left(\frac{1}{\Omega}-a\sin\theta\right)^2+r\sin^2\theta \ , \nonumber\\
   \beta_2&=&\sin\theta\cos\theta
   \left[\Delta+\frac{2r}{\Sigma^2}
   \left(\frac{a}{\Omega}-(r^2+a^2)\right)^2\right]\ , 
\end{eqnarray}
and $d r_{\rm surf}/d \lambda$ and $d \theta_{\rm surf}/d \lambda$  
   determine the intersection of the isobaric surfaces  
   and the $(r,\theta)$ plane.  With the Boyer-Lindquist variables defined as $\Sigma = r^2 +a^2\cos^2\theta$ and $\Delta = r^2 +a^2 - 2r$, and we have normalised the black hole mass to be 1 so all lengths are in terms of gravitational radii, $\RG$.
   
This coupled set of differential equations may then be solved numerically to obtain the isobaric surfaces of the torus parametrised in terms of the affine parameter $\lambda$.
In particular the isobaric surface that describes the `surface' of the optically thick torus may be obtained by integrating away from the marginally stable orbit on its equator (Fuerst \& Wu 2004).

\section{Velocity Law}
\begin{figure}
\vspace{-1.7cm}\hspace{-1.5cm}\epsfig{figure=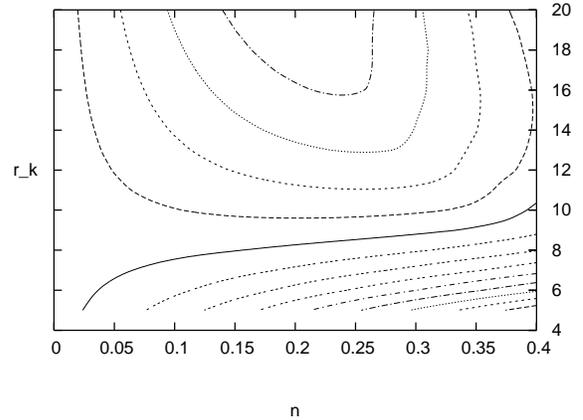,width=12cm}
\vspace{-1cm}
\caption{
Contours of likelihood for fitting equation \Eqn{rk_equation} to the velocity profile given in figure 13 of De Villiers and Hawley (2003) for their SFR model, corresponding to $a=0.998$.  This fit is truncated at $r_{\rm k}=20$ since the source graph ends there.  We have also ignored the points inwards of $3\RG$ where the gases angular momentum profile changes from being a rough power-law as it falls into the black hole.  For these assumptions, the best fit value of $n$ is around 0.2.  If points beyond of $20\RG$ are included the best fit value of $n$ slowly decreases towards zero as the torus becomes more Keplerian in character.
We use the above combined with the surface-finding method described in Fuerst \& Wu (2004) and Fuerst (2005) to generate the tori in this paper.
}
\label{contour_plot}
\end{figure} 

The above method requires a model of $\omega$, the angular velocity as a function of position.
We assume that it may be described by a function with the form
\begin{equation}
  \Omega=\frac{1}{(r\sin\theta)^{3/2}+a}
  \left(\frac{r_{\rm k}}{r\sin\theta}\right)^n \  ,  
\label{rk_equation} 
\end{equation}
where $r_{\rm k}$ is the radius (on the equatorial plane) 
   at which the material moves with a Keplerian velocity. 
The index  $n$ is a parameter to be determined below.
This form was chosen because tori require material to be flowing with faster than Keplerian velocity inside some point $r_{\rm k}$ on the equator, and slower than Keplerian velocity outside of it.  The differing rotational speed requires implicit pressure forces to exist, which in turn support the torus out of the equatorial plane.   Note that the velocity law chosen is a function of $r \sin\theta$.  In the Newtonian limit this relation causes the iso-density and isobaric surfaces to coincide.  This allows the assumption of a polytropic equation of state for the torus material (see Frank, King and Raine 1985).

Accretion tori are globally unstable to non-axisymmetric perturbations 
  (Papaloizou \& Pringle 1984;  Kojima 1986). 
The instabilities are consequences of interactions of waves on the inner and outer torus edges, 
   and the unstable modes grow on dynamical timescales.  
However, with the presence of a non-negligible radial inflow accretion component, 
   the reflective inner boundary is lost, 
   and the instabilities can be greatly suppressed 
   (Blaes 1987; Hawley 1991; Gat \& Livo 1992).  
As the exact stability properties of parametric model tori depend 
   on the assumed velocity (or angular momentum) profiles, 
   we may use the stability criteria to constrain the parameters of the velocity law 
   and hence the aspect ratio of parametric tori.   

\begin{figure}
\vspace*{0.3cm}
\center{\epsfig{figure=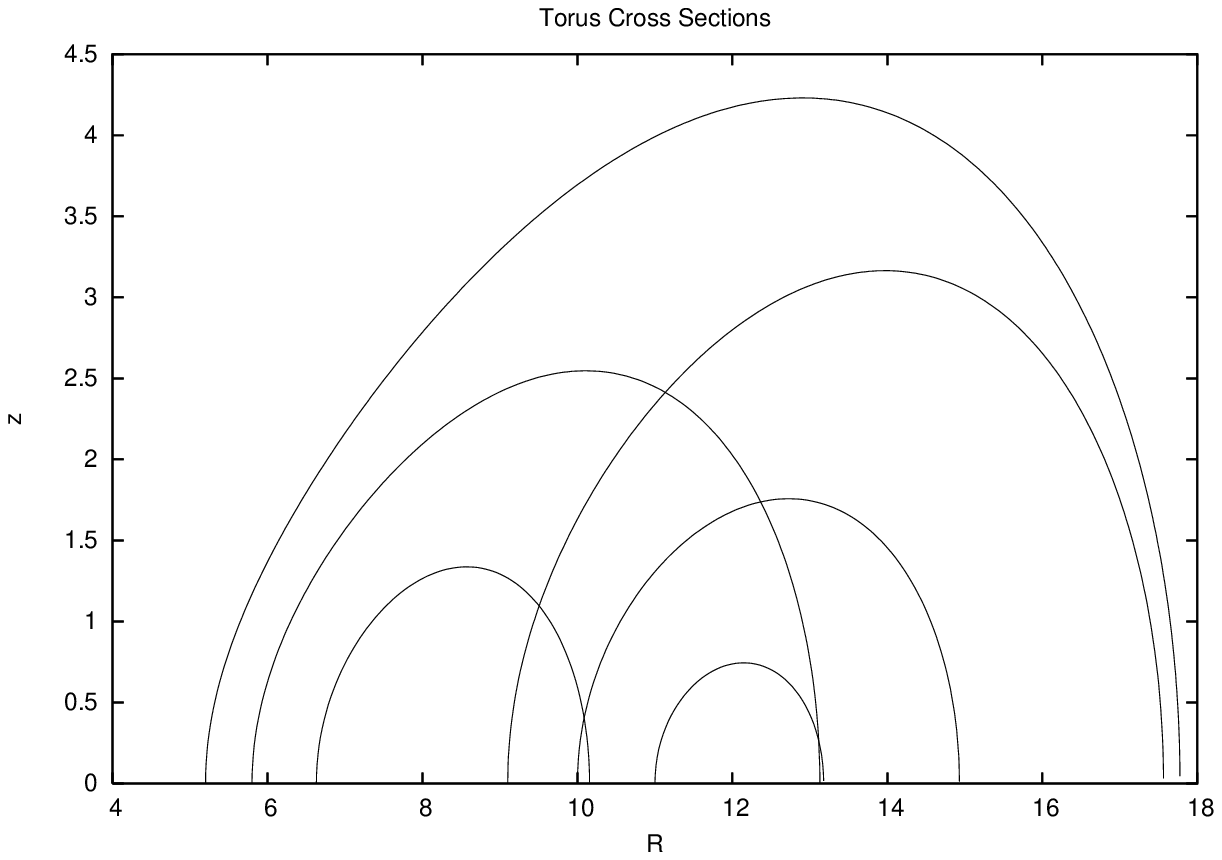,width=9cm}
\tiny
\begin{tabular}{|c|cccccc|}
\hline
$n$&0.15&0.18&0.232&0.232&0.232&0.232\\
$r_{\rm k}$&8&8&8&12&12&12\\
$a$&0.998&0.998&0.998&0.998&0.5&0\\
$r_{\rm in}$&5.1905&5.7670&6.6263&9.0585&9.9572&10.9053\\
$r_{\rm out}$&17.8929&13.3098&10.1644&17.7176&15.0218&13.3126\\
\hline
\end{tabular}}
\caption{
Table of parameters used for the various torus models in this paper together with the inner and outer radii of the tori.  The graph above shows the geometrical shape of the tori in cylindrical coordinates, with $R=r\sin\theta, z=r\cos\theta$.}
\label{table_size}

\end{figure}
\begin{figure}
\vspace*{0.3cm}
\epsfig{figure=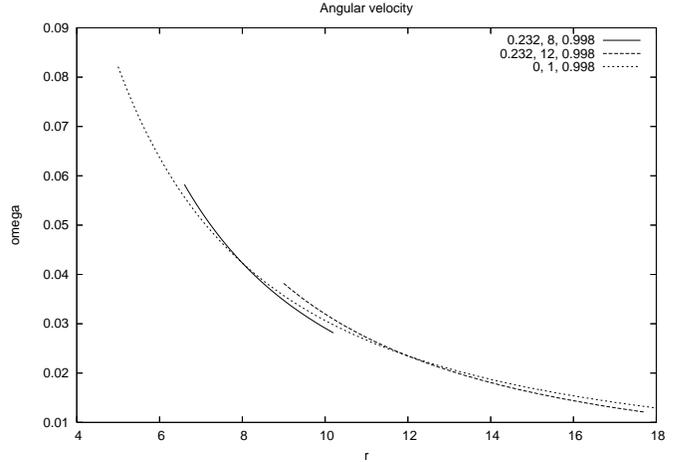,width=9cm}
\caption{
The angular velocity distribution of two tori in the equatorial plane, with $a=0.998$, $n=0.232$ and $r_{\rm k}=8$ (solid) or 12 (dashed).  The angular velocity distribution of a Keplerian disk (dotted) is also plotted here for comparison.  The angular velocity profile of the tori is sub-Keplerian at large radii, and super-Keplerian at small radii.  $r_{\rm k}$ is the most important parameter in determining $\omega(r)$.  Changing $n$ and $a$ only weakly affects the shape of this function, and so are not altered here as the graphs would lie on top of each other.
}
\label{velocity_plot}
\end{figure}

The parameter $n$ in equation \ref{rk_equation}. is roughly related to the $q$ index of the von Zeipel parameter 
   (see e.g. Chakrabarti 1985; Blaes \& Hawley 1988),    
   which is often used in the instability study of accretion disks,   
   via $n \approx q - 1.5$.  The relation is exact when the black hole spin parameter $a=0$.
Analyses show that Newtonian tori with $q>\sqrt{3}$ are generally unstable.        
Fitting the profiles of relativistic tori obtained 
    by numerical magnetohydrodynamic simulations of De Villiers \& Hawley (2004)  
    yields $n \approx 0.2$, corresponding to $q \approx 1.7$.  (See Fig. \ref{contour_plot}.)
In general, fatter tori have smaller values of $n$. 
However, the aspect ratio of a torus depend only weakly on $n$.  

The shapes of the tori modelled in this paper are shown in Figure~\ref{table_size}, 
where the radii of the inner and outer orbits are also tabulated.  
Figure \ref{velocity_plot} shows the angular velocity $\Omega$ as a function of radius in the equatorial plane.  Note that unless otherwise stated, we consider model tori with $n = 0.232 \simeq \sqrt{3}-1.5$ in Kerr space-time with a spin parameter $a=0.998$.

We calculate line profiles by integrating the flux over the ray-traced images of the tori.  
As in Fuerst \& Wu (2004), the line emissivity is assumed to be a power law with ${\cal I} \propto r^{-k}$.  
For the ray tracing, torus images of $1000\times1000$ pixels are constructed.  
The red shift of the emission corresponding to each pixel is calculated. 
The line profiles are obtained by summing the flux, 
  and 100 energy bins are used for the line spectra.

\begin{figure} 
\vspace*{0.5cm}
\center{\epsfig{figure=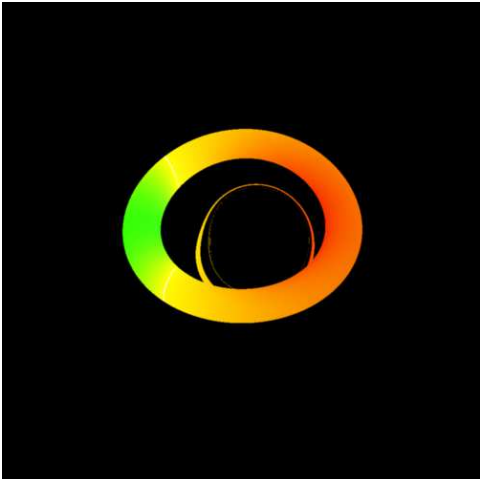,width=6.35cm}}
\center{\epsfig{figure=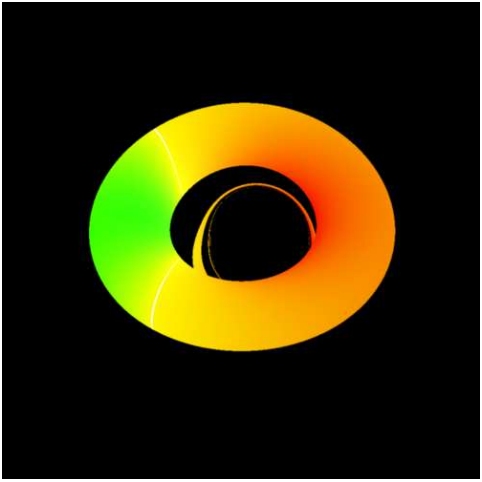,width=6.35cm}}
\center{\epsfig{figure=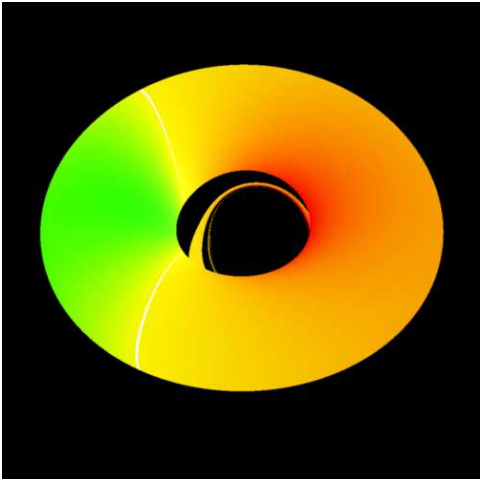,width=6.35cm}}

\caption{ 
   Energy shifts (with respect to a distant observer)   
      at the boundary surfaces of accretion tori with aspect ratios 
      corresponding to rotational velocity power-law indices 
      $n = 0.232$, 0.18 and 0.15 (panels from top to bottom respectively). 
   The spin parameter of the central black hole is  $a = 0.998$,   
      the radius of Keplerian rotation is $r_{\rm k} = 8\RG$, and 
      the viewing inclination angle of the tori is $i = 45^\circ$. 
   See Fig.~\ref{image_disk} for a description of the colour scale.
 }
\label{image_n}
\end{figure} 

\begin{figure} 
\vspace*{0.5cm}
\center{\epsfig{figure=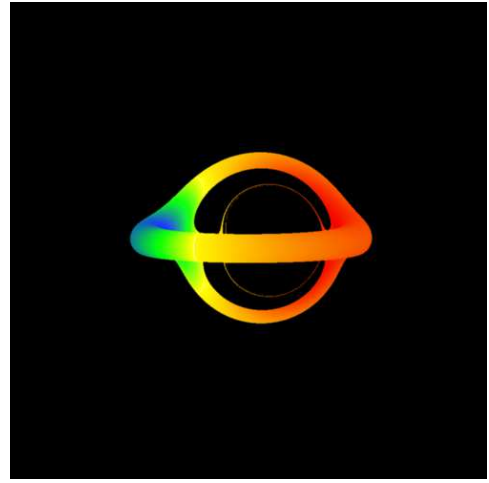,width=6.35cm}}
\center{\epsfig{figure=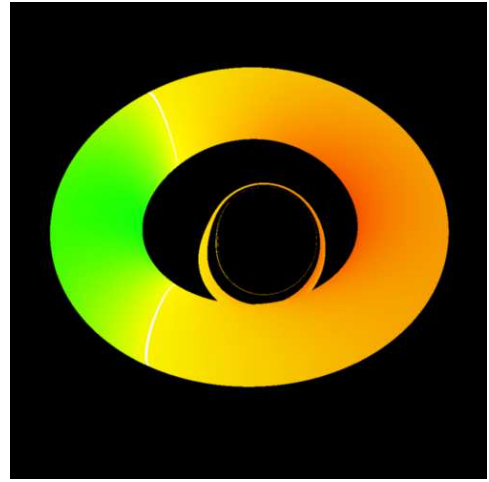,width=6.35cm}}
\center{\epsfig{figure=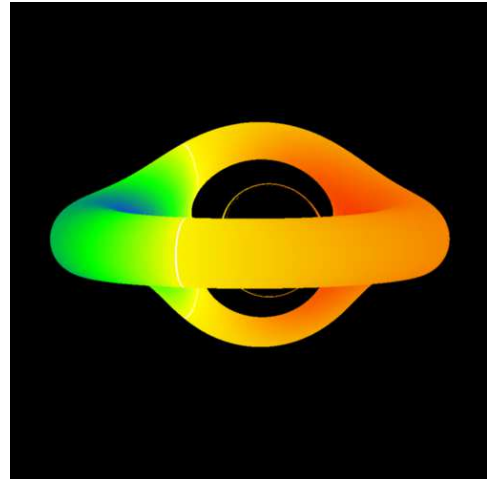,width=6.35cm}}

\caption{
 (Top): Same as top panel of Fig.~\ref{image_n}, i.e. $r_{\rm k} = 8\RG$, 
       but viewed at an inclination angle of $85^\circ$.  
  (Middle): Energy shifts (with respect to a distant observer)  
       at the boundary surface of an accretion torus 
       with rotational velocity power-law index $n = 0.232$ 
       and Keplerian radius $r_{\rm k} = 12\RG$. 
       The viewing inclination angle is $i=45^\circ$.    
       The other parameters are the same as those of the tori in Fig.~\ref{image_n}.  
  (Bottom): Same as middle panel but viewed at an inclination angle of $85^\circ$.  }
\label{image_4585rk}
\end{figure}

This technique has several limitations in representing the results of GRMHD simulations.  Firstly, the analytic velocity distribution used here does not contain any component in the $\hat{r}$ or $\hat{\theta}$ directions thus the assumed inflow used to derive the torus stability, and hence its shape, is not included.  However, since the turbulence and inflow is subsonic, and the disk bulk motion is supersonic, this approximation only slightly narrows the resulting line profiles.

The model here also only investigates the inner part of the accretion disk and thus does not approach the behaviour of standard thin accretion disk models at large radii.  However, since it is known that the relativistic lines appear to be generated by material extremely close to the black hole via {\em Suzaku} observations of MCG-6-30-15 (Miniutti \etal 2006), the neglect of line emission from the thin disk is not important.

In common with most other analytic models of accretion disks, the model explored here is time independent.  In general, MHD simulations of accretion disks are highly turbulent and show extreme variability (Hirose \etal 2006).  Thus, the results explored here must be seen as a time-average of the true properties of these objects.

\section{Results and discussion}  

\begin{figure} 
\vspace*{0.3cm}
\center{\epsfig{figure=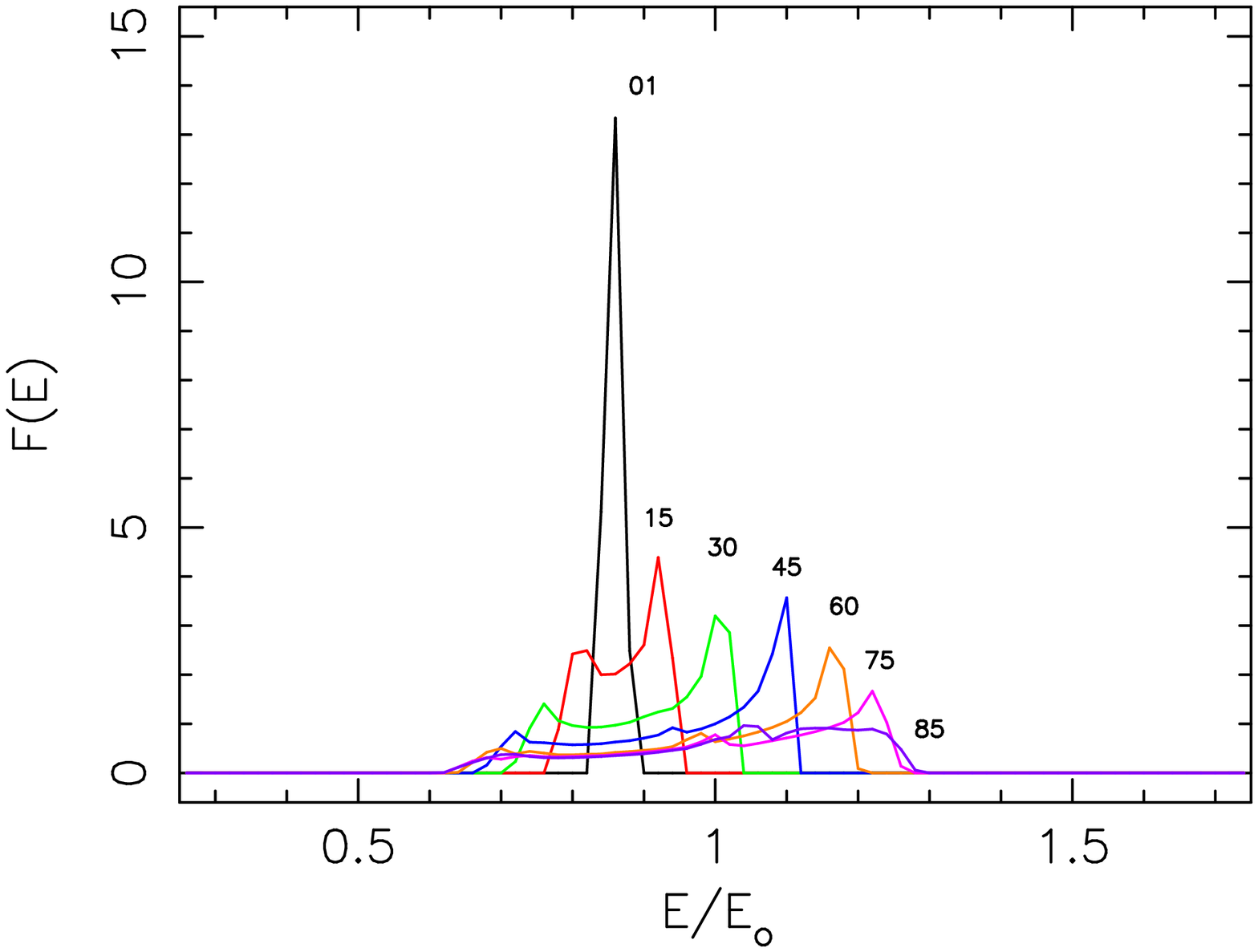,width=7.75cm}} 
\center{\epsfig{figure=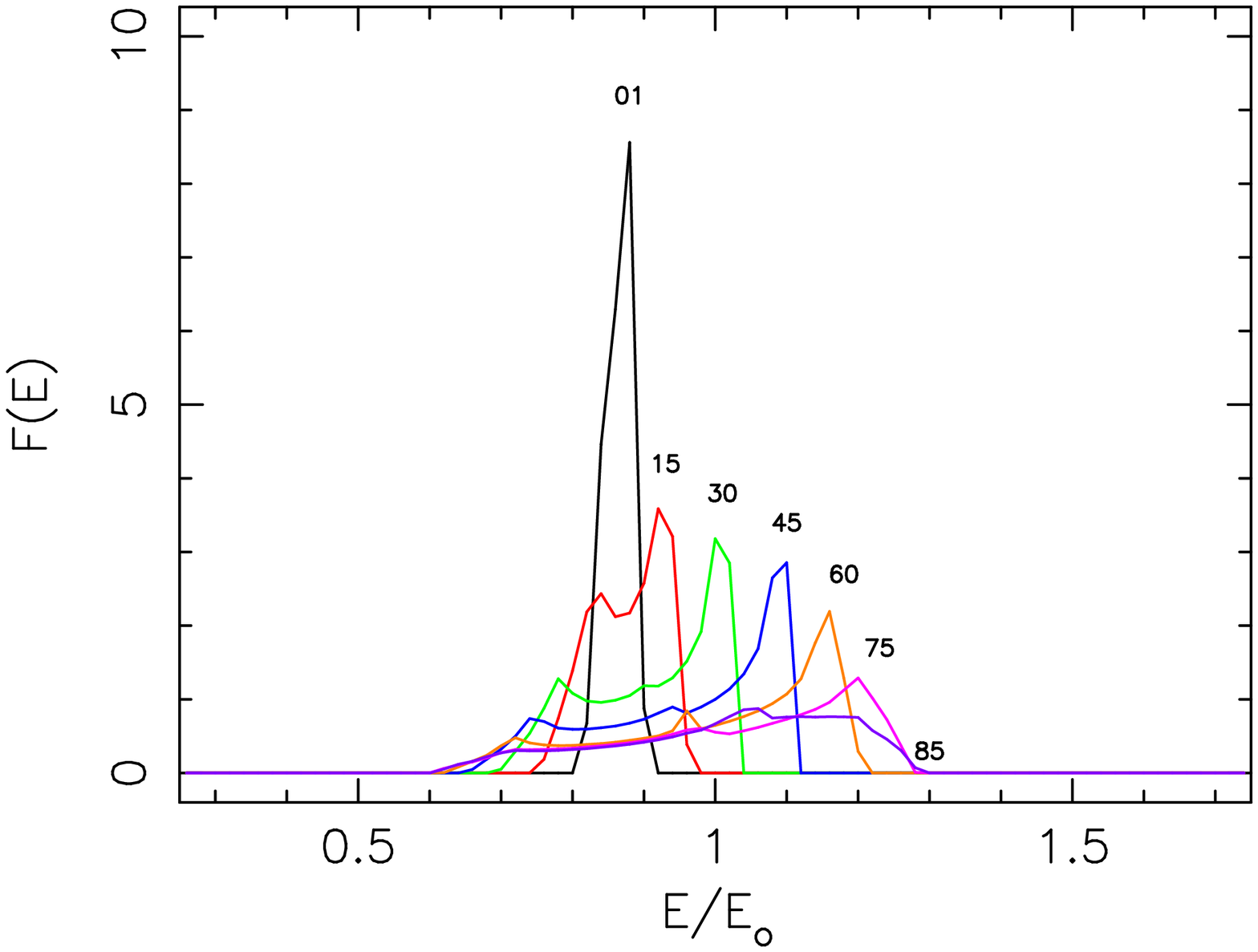,width=7.75cm}}
\center{\epsfig{figure=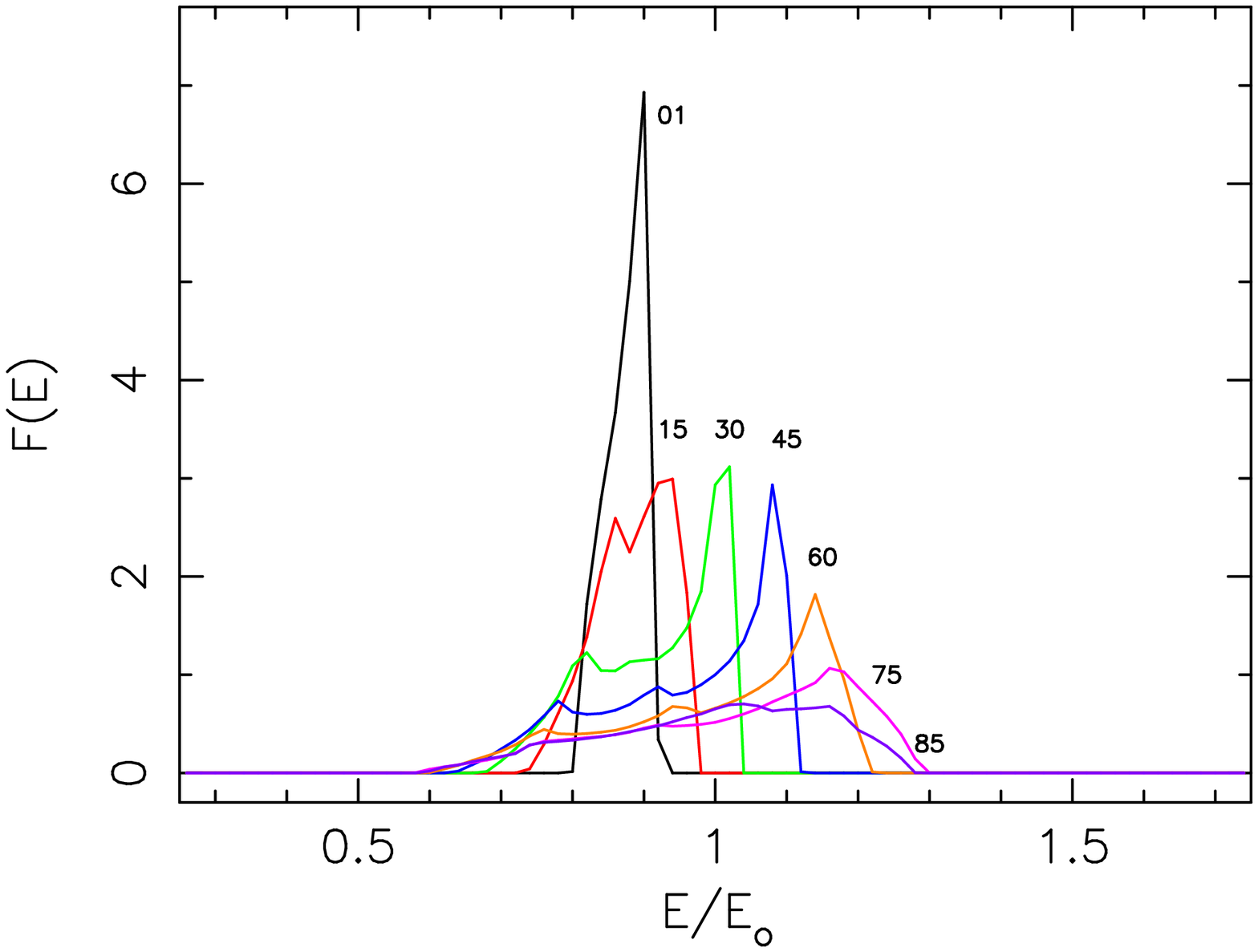,width=7.75cm}}
\caption{
  Profiles of emission lines from accretion tori viewed at inclination angles 
      $i = 1^\circ$,  $15^\circ$, $30^\circ$,  $45^\circ$, $60^\circ$, $75^\circ$ and  $85^\circ$.  
  The radial emissivity power-law index is $-2$.  
  The radius of Keplerian rotation is $r_{\rm k} = 12\RG$, 
      and the rotational velocity power law index $n=0.232$.  
  The spin parameters of the central black holes are
      $a=0$, 0.5 and  0.998 (panels from top to bottom).  
   The normalisation is such that $F(E) = 1$ at $E/E_{\rm o} = 1$  
       for $i=45^\circ$.     }
\label{v_spin}
\end{figure}

\begin{figure} 
\vspace*{0.3cm}
\center{\epsfig{figure=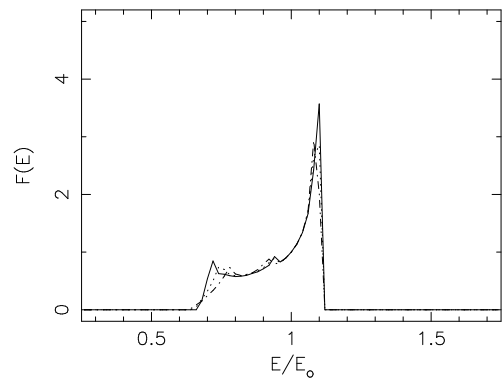,width=7.75cm}}
\caption{  Comparison of line profiles of accretion tori 
   for black hole spin parameters  $a = 0$, 0.5 and 0.998 
   (represented by solid line, dotted line and  dot-dashed line respectively).   
   The  viewing inclination angle is $i=45^\circ$.  
   The other parameters are the same as Fig. \ref{v_spin}. 
   The normalisation is such that $F(E) = 1$ at $E/E_{\rm o} = 1$. }
\label{v_a}
\end{figure}

\begin{figure} 
\vspace*{0.3cm}
\center{\epsfig{figure=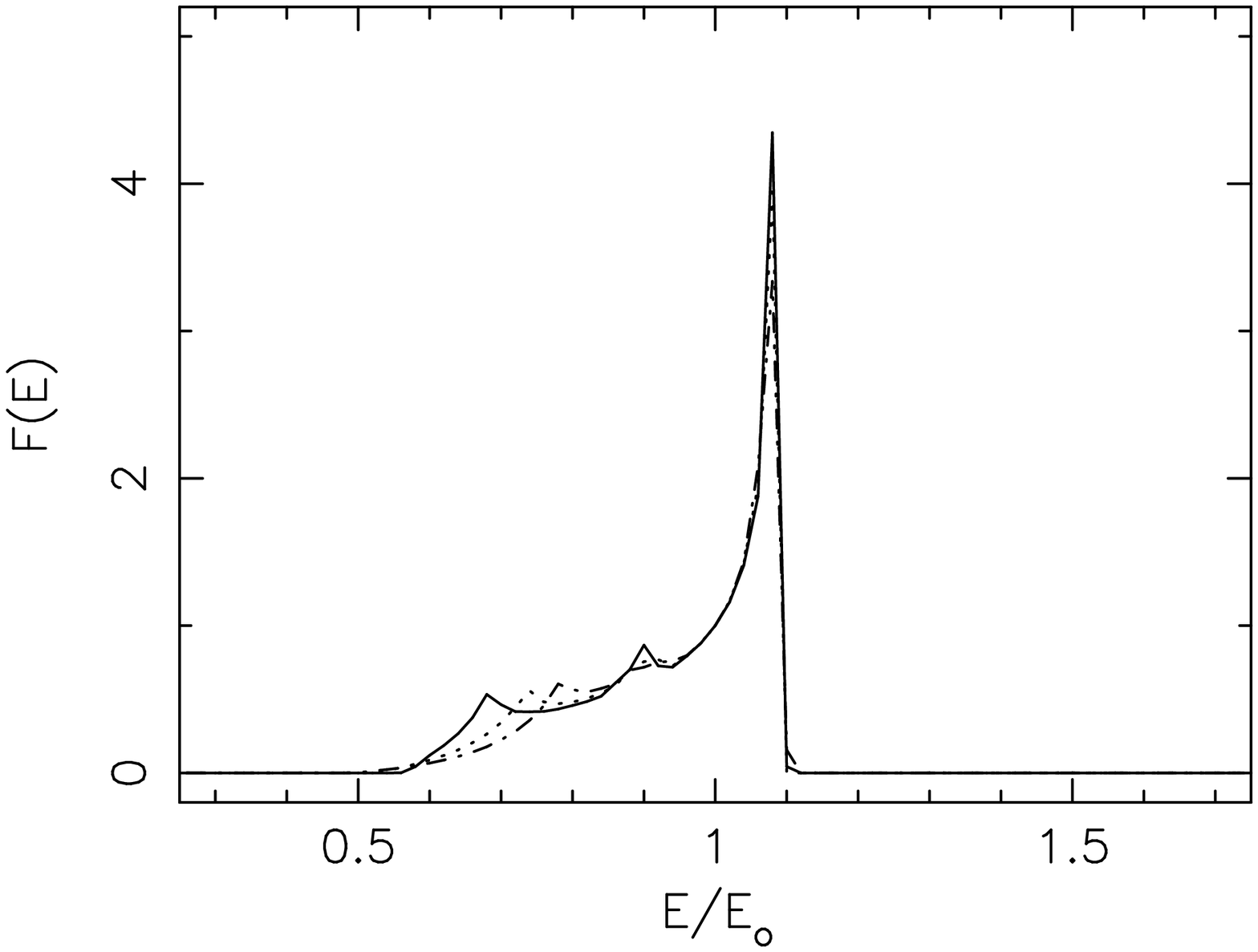,width=7.75cm}}
\center{\epsfig{figure=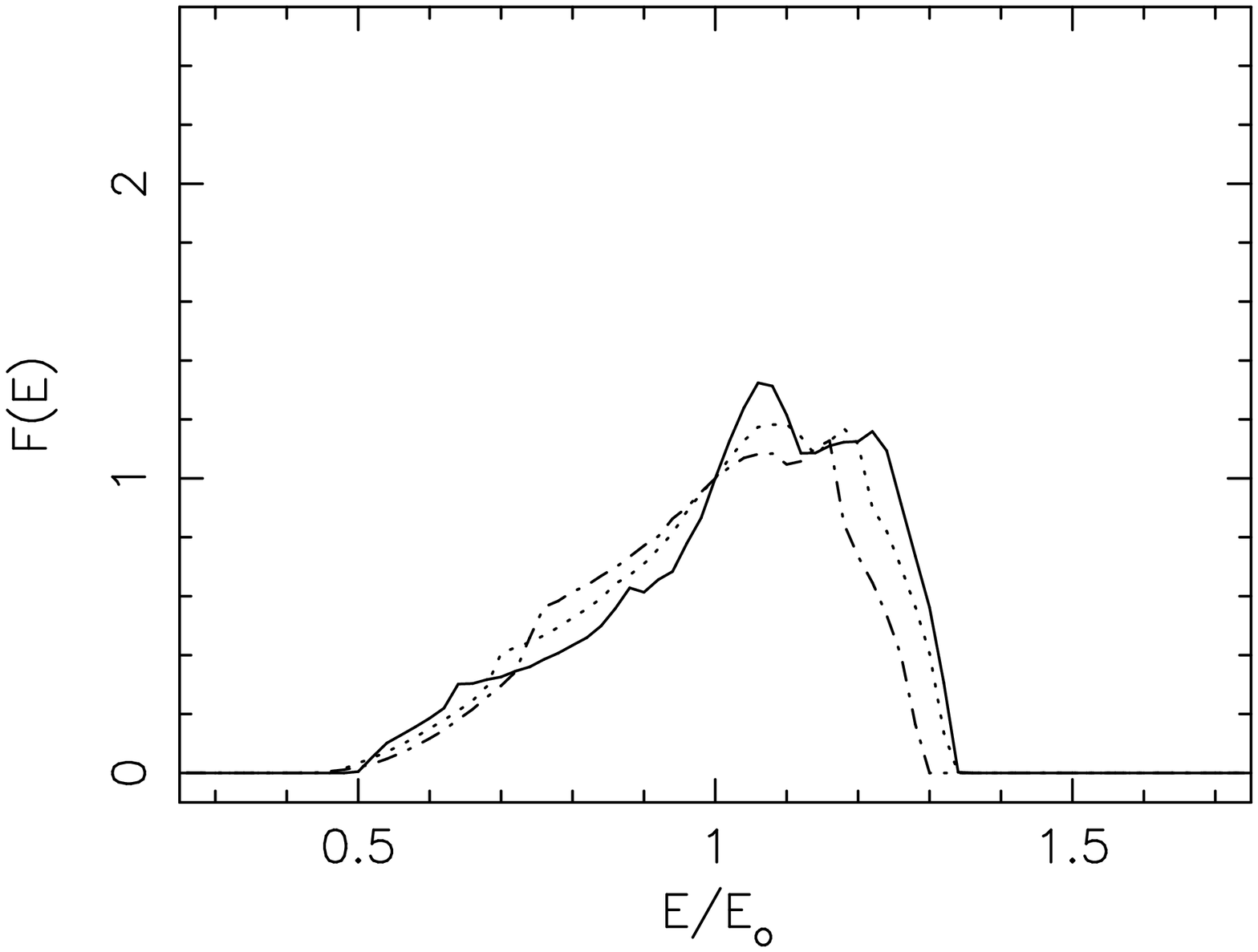,width=7.75cm}}
\label{v_n}
\caption{
  Profiles of lines from accretion tori with aspect ratios  
     corresponding to rotation velocity power-law indices   
     $n=0.232$, 0.18 and 0.15 
     (represented by solid line, dotted line and dot-dashed line respectively). 
 The spin parameter of the central black hole is $a=0.998$  
    and the radius of Keplerian rotation is $r_{\rm k}=8\RG$.      
 The tori have emissivity power-law index of $-2$.  
 The viewing inclination angles of the tori are  $45^\circ$ (top panel) 
      and $85^\circ$ (bottom panel).  
  The normalisation is such that the flux $F(E) = 1$ at $E/E_{\rm o} = 1$, 
      where $E_{\rm o}$ is the line centre energy in the rest-frame.  }
\label{line_4585}
\end{figure}

What role does geometry play in determining the line emissions from accretion tori?  
There are two aspects of geometrical effects: 
   one concerning the intrinsic geometry of the torus, 
   and another one concerning the viewing inclination of the system.  
In the accretion torus model considered here, 
   the shape (aspect ratio) of a torus is determined by the rotational velocity power-law index  $n$, 
   and the linear extension of the torus is set mainly by the Keplerian radius $r_{\rm k}$. 
For fixed $r_{\rm k}$, a larger $n$ gives a more inflated torus (see Fig.~\ref{image_n});  
   for fixed $n$, a larger $r_{\rm k}$ gives a larger outer torus radius 
   (cf. tori in Fig.~\ref{image_n} and \ref{image_4585rk}).  
The effects of viewing inclination are more complicated.  
Clearly, the projection area of the visible regions of a torus onto the sky plane 
   depends on the viewing inclination angle $i$.    
The torus, which has considerable thickness, can be self-eclipsed. 
It can also be gravitational lensed by the central black hole.  
Lensing effects are important especially when $i$ is large.   
In extreme situations ($i \approx 90^\circ$),  
   the projection areas of the lensed bottom parts of the tori 
   are comparable with the areas of their upper parts (see Fig.~\ref{image_4585rk}, top and bottom panels).    
See Viergutz (1993) for wire-frame images of tori, and Bursa \etal (2004) 
   for images of semi-transparent tori around Schwarzschild black holes.
   
The images of the  accretion tori in Figures \ref{image_n} and \ref{image_4585rk} 
    are asymmetric and there is a series of thin rings in the ``hole'' of each projected torus image. 
This is due to the fact that the central black holes of the tori are very fast spinning (with $a = 0.998$). 
The left-right symmetry is destroyed because of reference-frame dragging by the rotation of the black hole.  
The rings are produced by lensed photons orbiting the black hole.  
Each ring corresponds to a family of indirect photon paths from the torus surface to the observer.   
There are an infinite number of these rings, corresponding to an infinite number of image orders.     
However, their contribution to the total emission decreases rapidly with the image order, 
   and photons with more than two black hole orbits reaching the observer are of insignificant number.  
(See Beckwith and Done (2005) for a discussion of these higher order images from thin disks.) 
  
\subsection{Line profiles}  
    
We now analyze how the line profiles depend on 
   viewing inclination, black hole spin, aspect ratio and linear extension of the torus, 
   and spatial emissivity profile on the torus surface.  
Figure~\ref{v_spin} shows the emission lines from accretion tori  
   with $n = 0.232$ and $r_{\rm k} = 12\RG$  
   around black holes with spin parameters $a= 0$, 0.5 and 0.998 (panels from top to bottom) 
   at various viewing inclination angles.  
The lines are single-peaked for small $i$ (see the line profiles corresponding to $i =1^\circ$).   
Moreover, the line centroids are severely redshifted.     
As $i$ increases, the lines are broadened and the line centroids migrate blueward.  
At the same time, a sharp blue line peak begins to develop. 
For sufficiently large $i$, the red and blue line peaks become clearly distinguishable, 
   and the lines resemble those of geometrically thin accretion disks. 
As $i$ increases further (approaching $90^\circ$), the blue line peak is gradually suppressed. 
However, another peak begins to emerge. 
This peak is due to high-order lensed emissions. 
It is weak and has a small energy redshift. 
For $i$ close to $90^\circ$, the line peaks are not very clearly distinguishable, 
   and the lines appear to be broad, asymmetric and single-peaked.   
The trend of line morphology changing with $i$ is similar 
   for all the black hole spin parameters $a$.  
  
When the tori are viewed almost pole on, 
   the main difference between lines from tori 
   around a Schwarzschild ($a = 0$) and a maximally rotating Kerr black hole ($a=0.998$) 
   is that the line of the latter is broader and more asymmetric. 
The redshifts of the line centroids are similar. 
For moderate viewing inclination angles ($i \sim 45^\circ$),   
   the redshift of the red line peak becomes smaller when $a$ increases.   
On the one hand the red wing of the line is suppressed, while 
  on the other hand, the maximum redshift of the red line wing increases.
The line profile might be narrower for larger $a$.  
The situation is more complicated for $i$ close to $90^\circ$, 
   because of various competing factors which are difficult to disentangle.   
    
The apparent weak dependence of the line profiles on the black hole spin can be attributed to the following. 
Firstly, the inner boundary surface of the torus at the equatorial plane 
   is the innermost stable orbit (see Fuerst \& Wu 2004),
   which is determined by the balance between gravitational and pressure forces. 
Unless $r_{\rm k}$ is very small, the torus inner boundary is not close to the black hole event horizon.    
As the dynamics of accretion flow and thus the shape of the emission region are not greatly affected by the black hole spin,  
   the integrated emission from the torus is insensitive to this parameter.      
Secondly, at large inclination angles, the accretion tori self-eclipse. 
When self-eclipse occurs, 
   the most highly redshifted and blueshifted emission from the inner torus regions are blocked from view, 
   and the emission is mostly contributed by the outer torus surface. 
The eclipsing process is determined by the viewing angle and torus aspect ratio, which is practically independent of $a$.  
Thirdly, although the black hole spin can greatly affect gravitational lensing, 
  the contributions of high-order lensed emissions to the total emission are small.    
  
Figure~\ref{line_4585}  shows the profiles of emission lines from three tori with aspect ratios 
    corresponding to velocity power-law indices $n = 0.232$, 0.18 and 0.15. 
At $i= 45^\circ$, the emission lines have asymmetric double-peak profiles.  
The location of the blue line peak is roughly the same for tori with different aspect ratio,     
   but the relative height of the blue peak decreases with $n$. 
The red peak changes with the aspect ratio of the torus.  
The peaks of tori with smaller $n$ have smaller redshifts. 
They also have weaker red wings.
At $i = 85^\circ$, the lines are broad, asymmetric and single-peaked. 
They also have an extensive red wing. 
Both the red and blue line wings change with the aspect ratio of the torus, 
  and lines with smaller $n$ are in general narrower. 

The dependence of line profiles on the aspect ratios of the tori can be understood as follows.  
At small or moderate viewing inclination angles,  
    the projected area of the visible surface of a torus on the sky plane  
    increases with the flatness of the torus (see Fig.~\ref{image_n}).   
Emissions from the inner torus regions 
   are the most relativisticly boosted or the most gravitationally redshifted.       
As the size of a torus increases,  
   the relative contribution of the emission by the inner torus regions decreases, 
   hence reducing the height of the blue line peak and suppressing the red line wing.  
At sufficiently large viewing inclination angles, 
   occultation and lensing become more important.  
Self-eclipse blocks the Doppler boosted blueshifted emissions  
   and the gravitationally redshifted emissions.    
At the same time, lensing brings the bottom part of the torus into view.  
The emissions from the newly visible lensed regions are not strongly relativistically boosted 
   but are slightly redshifted because of time dilation  
   (which is due to the transverse motion of the emitters 
   and the gravity of the central black hole).  

\begin{figure} 
\vspace*{0.3cm}
\center{\epsfig{figure=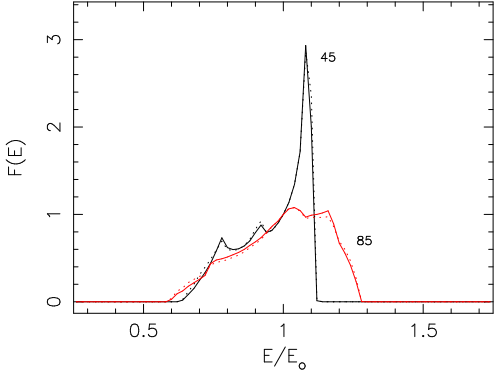,width=7.75cm}}
\caption{  Comparison of line profiles for different emissivity power-law indices 
    at viewing inclination angles $i=45^\circ$ and $85^\circ$. 
  Lines for emissivity power-law index $-2$ is represented by solid lines and   
    lines for emissivity power-law index $-3$ is represented by dotted lines. 
  The spin parameter of the central black hole is $a=0.998$. 
  The other parameters are the same as those for tori in Fig. \ref{v_spin}. 
  The normalisation is such that $F(E) = 1$ at $E/E_{\rm o} = 1$. }
\label{v_g}
\end{figure}
  
\begin{figure} 
\vspace*{0.3cm}
\center{\epsfig{figure=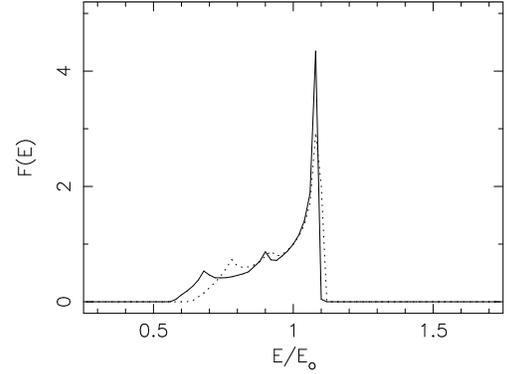,width=7.75cm}} 
\center{\epsfig{figure=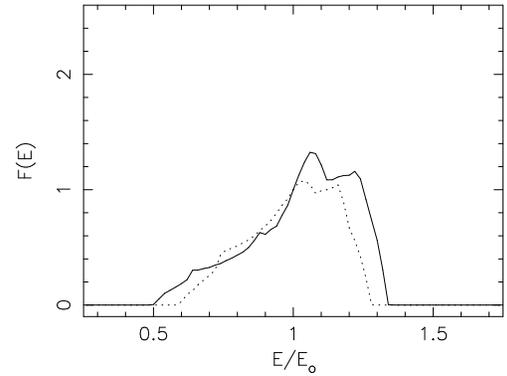,width=7.75cm}}
\caption{ 
  Profiles of lines from accretion tori with $r_{\rm k} = 8\RG$ (solid line) and $12\RG$ (dotted line). 
  The spin parameter of the central black hole is $a=0.998$, 
     the rotational velocity  power law index is $n=0.232$, and 
     the the radial emissivity power law index is $-2$.  
  The viewing inclination angles of the tori are $45^\circ$ (top panel)  
      and $85^\circ$ (bottom panel).  
  The normalisation is such that $F(E) = 1$ at $E/E_{\rm o} = 1$. }
\label{v_rk}
\end{figure}

Figure~\ref{v_g} shows a comparison of two tori with the same parameters 
   except the emissivity power-law index (of values $-2$ and $-3$) 
   viewed at $i= 45^\circ$ and $85^\circ$. 
Despite that line profiles change with $i$,   
   the lines of the two tori are almost identical.  
The effects of altering the radial emissivity power-law index are small   
   due to the fact that the difference between the outer and inner boundaries of the torus 
   is relatively small,
   and that the innermost orbit of the emitters is quite far from the black hole event horizon. 

Figure~\ref{v_rk} shows the profiles of lines from accretion tori 
   with linear extensions given by $r_{\rm k} = 8\RG$ and $12\RG$. 
At $i = 45^\circ$, the lines from both tori have asymmetric profiles and three peaks 
    --- the usual red and blue peaks for accretion disks/tori 
         and in addition a small central peak 
         corresponding to the high-order lensed emissions. 
The locations of the line peaks of the tori are not the same. 
The red peak is located at a lower energy for the torus with $r_{\rm k} = 8\RG$. 
Also, the blue peak is narrower,  the blue edge at a lower energy,   
  and the central peak is at a slightly lower energy for this torus.  
Overall, the line from the smaller torus is ``redder''. 
At $i=85^\circ$, the line from  the torus with $r_{\rm k} = 8\RG$  
   is broader  than the line from the torus with $r_{\rm k} = 12\RG$. 
The red line wing is stronger for the smaller tori, as in the case of $i=45^\circ$. 
However, the blue edge is at a higher energy for the torus with $r_{\rm k} = 12\RG$, 
  which is  in contrast to the case of $i=45^\circ$.       

The inner boundary surfaces of the tori with $r_{\rm k} = 8\RG$ and $12\RG$ 
   are reasonably far from the event horizons of their central black holes.  
The inner radius of the former torus is roughly 2/3 of that of the latter torus, 
    similar to the ratio of their Keplerian radius $r_{\rm k}$ 
    (cf. the tori in the top panel of Fig.~\ref{image_n} 
    and in the middle panel of Fig.~\ref{image_4585rk}).     
The lower energy for the red line wing for the smaller torus when viewed at $i = 45^\circ$  
    is due to the lowest energy photons from it 
    being from deeper in the gravitational well of the black hole. 
The blue edge of the line is set by the photons with the highest energy. 
Its location is determined by 
   the magnitude of the relativistic Doppler blueshift of the emission 
   (due to line-of-sight motion of the emitters) 
   after being compensated for by gravitational redshift.   
The blue edge of the line from the torus with $r_{\rm k} = 8\RG$ 
   is at a higher energy than that from the torus with $r_{\rm k} = 12\RG$ 
   because the photons are emitted closer to the black hole from material moving
   at a higher velocity.
The kinematic Doppler shift outweighs the gravitational and transverse redshift in this particular case.
At very large viewing inclination angles, 
   self-eclipse and lensing are dominant effects. 
As shown in Figure~\ref{image_4585rk}, 
   the accretion torus with $r_{\rm k} = 8\RG$ is less eclipsed than 
   the torus with $r_{\rm k} = 12\RG$. 
The maximum blue shift and maximum red shift for the emissions 
   from the former torus are therefore larger, 
   as more of its inner regions are visible.  

\subsection{Comparison between lines from accretion tori and thin accretion disks}   

\begin{figure} 
\vspace*{0.3cm}
\center{\epsfig{figure=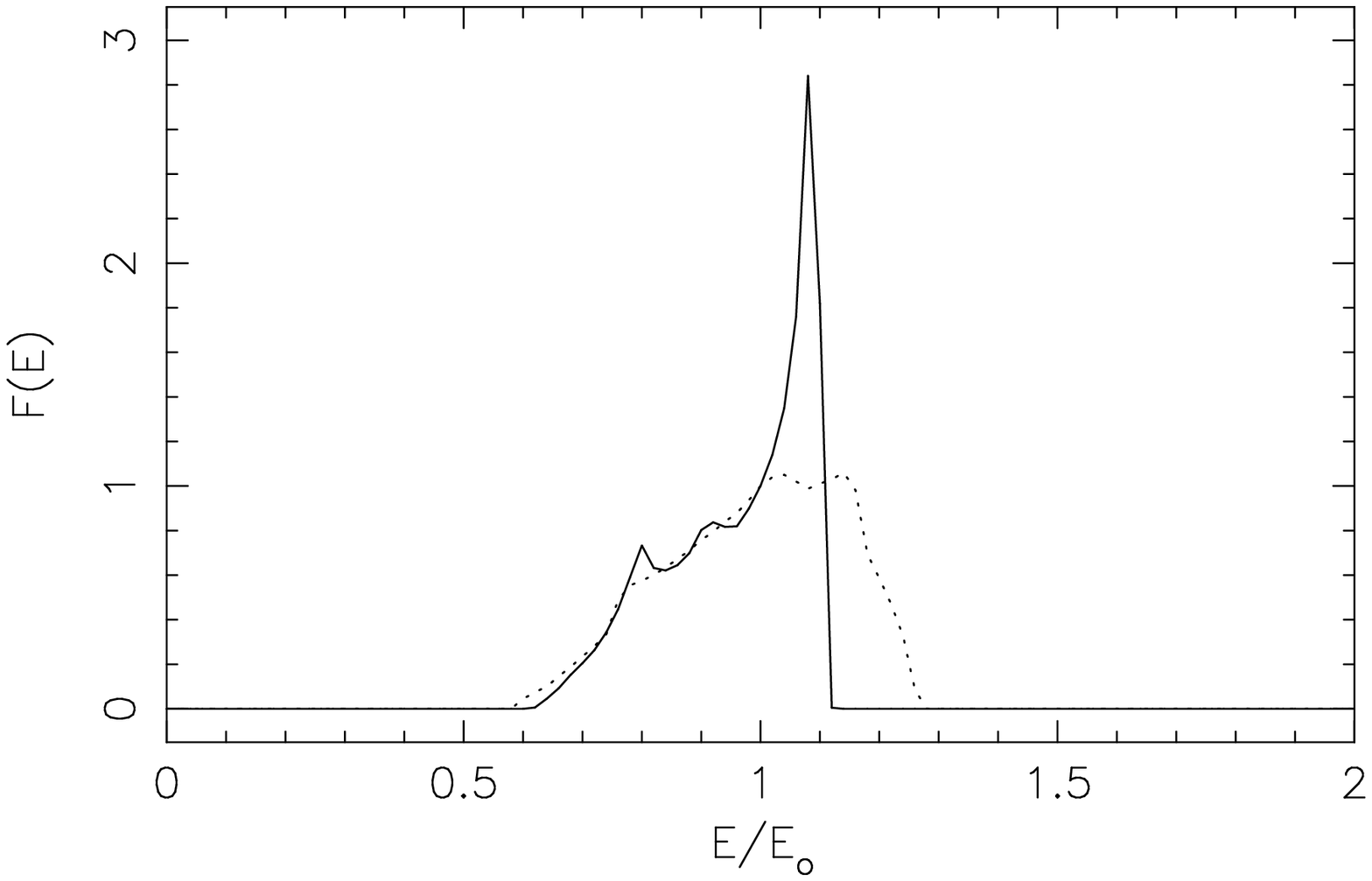,width=7.75cm}}
\center{\epsfig{figure=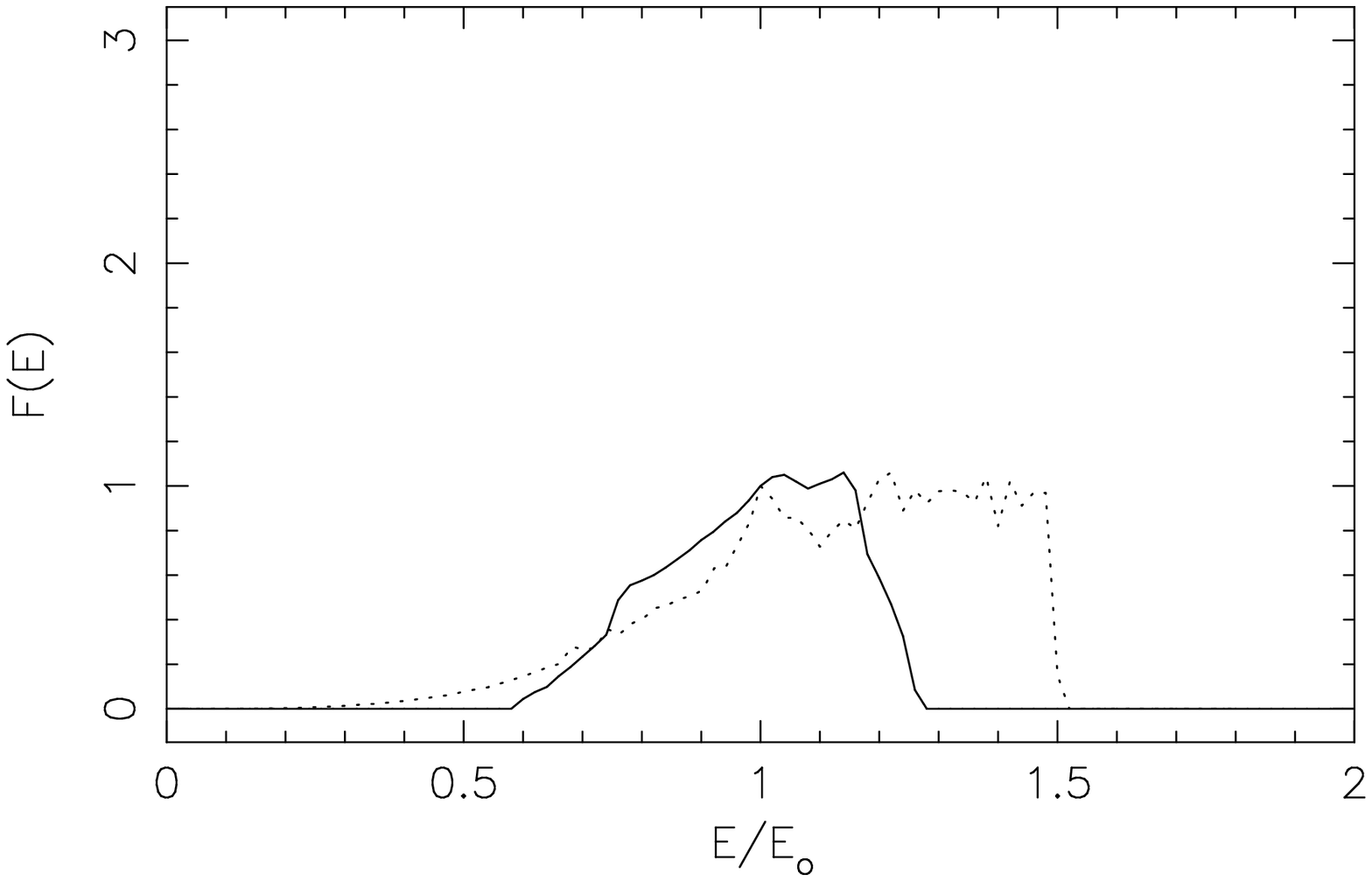,width=7.75cm}}
\caption{  Comparison of line profiles for  
    a geometrically thin accretion disk and an accretion torus 
    at viewing inclination angles $i=45^\circ$ and $85^\circ$
	around a black hole with spin parameter $a=0.998$.
  The parameters of the torus are $n=0.21$ and $r_{\rm k} = 12\RG$.
  This yields an inner marginally stable radius of $9.06\RG$,
   and an outer radius of $20.243\RG$.
  A torus with these parameters is shown in Fuerst \& Wu (2004),
  and this figure reproduces figures 6 and 7 of that paper to allow
  easier comparison with line profiles from those of the tori in this paper.
  The disk was chosen to have an inner radius at the marginally stable orbit,
  at $1.23\RG$ and an outer radius of $20\RG$.
  The line profiles for the accretion torus are represented by solid lines 
     and the line profiles for the accretion disk are represented by dotted lines.  
  The line emissivity power-law index is $-2$ for both the torus and the disk.
  The normalisation is such that $F(E) = 1$ at $E/E_{\rm o} = 1$ }
\label{old_dt}
\end{figure}

\begin{figure} 
\vspace*{0.5cm}
\center{\epsfig{figure=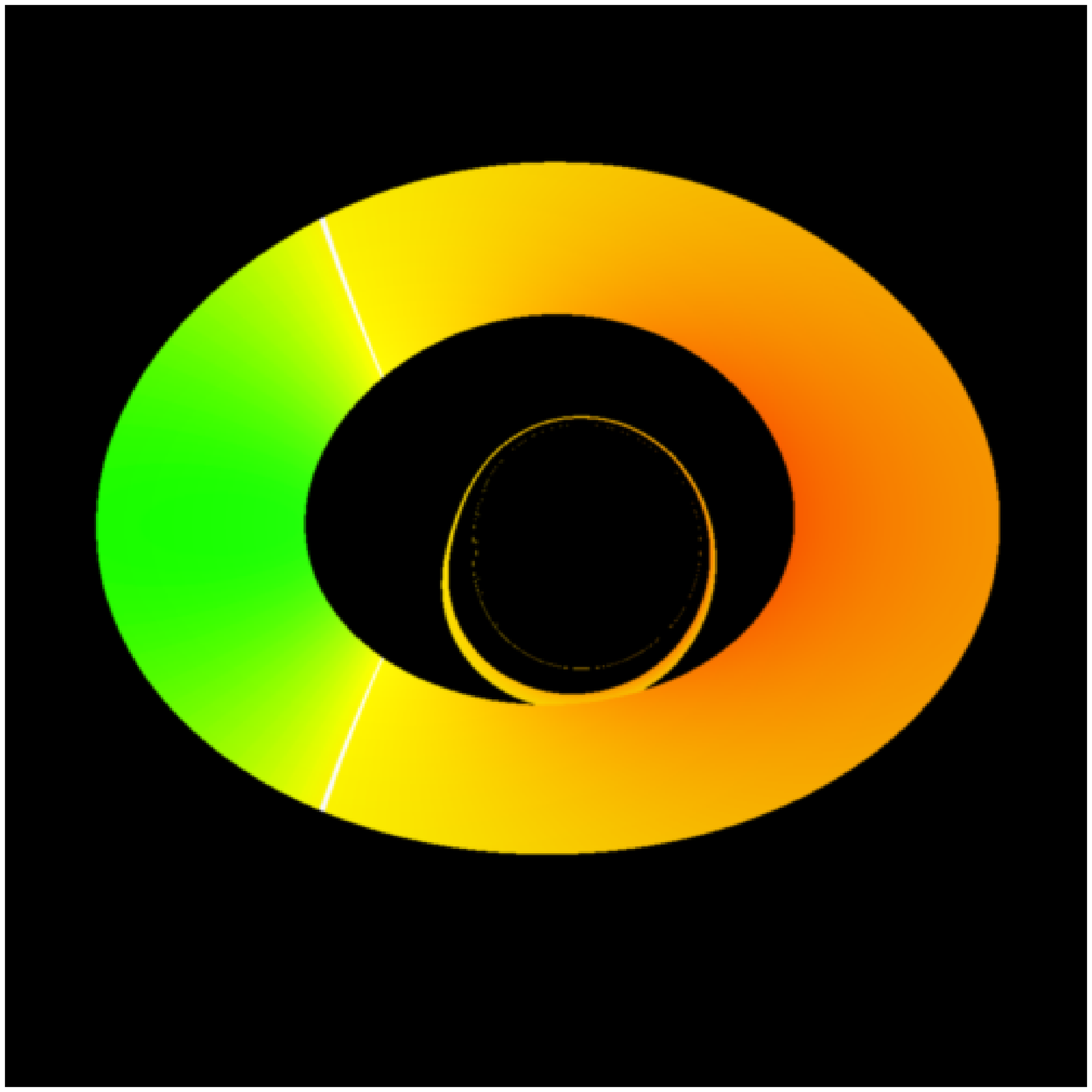,width=6.35cm}}
\center{\epsfig{figure=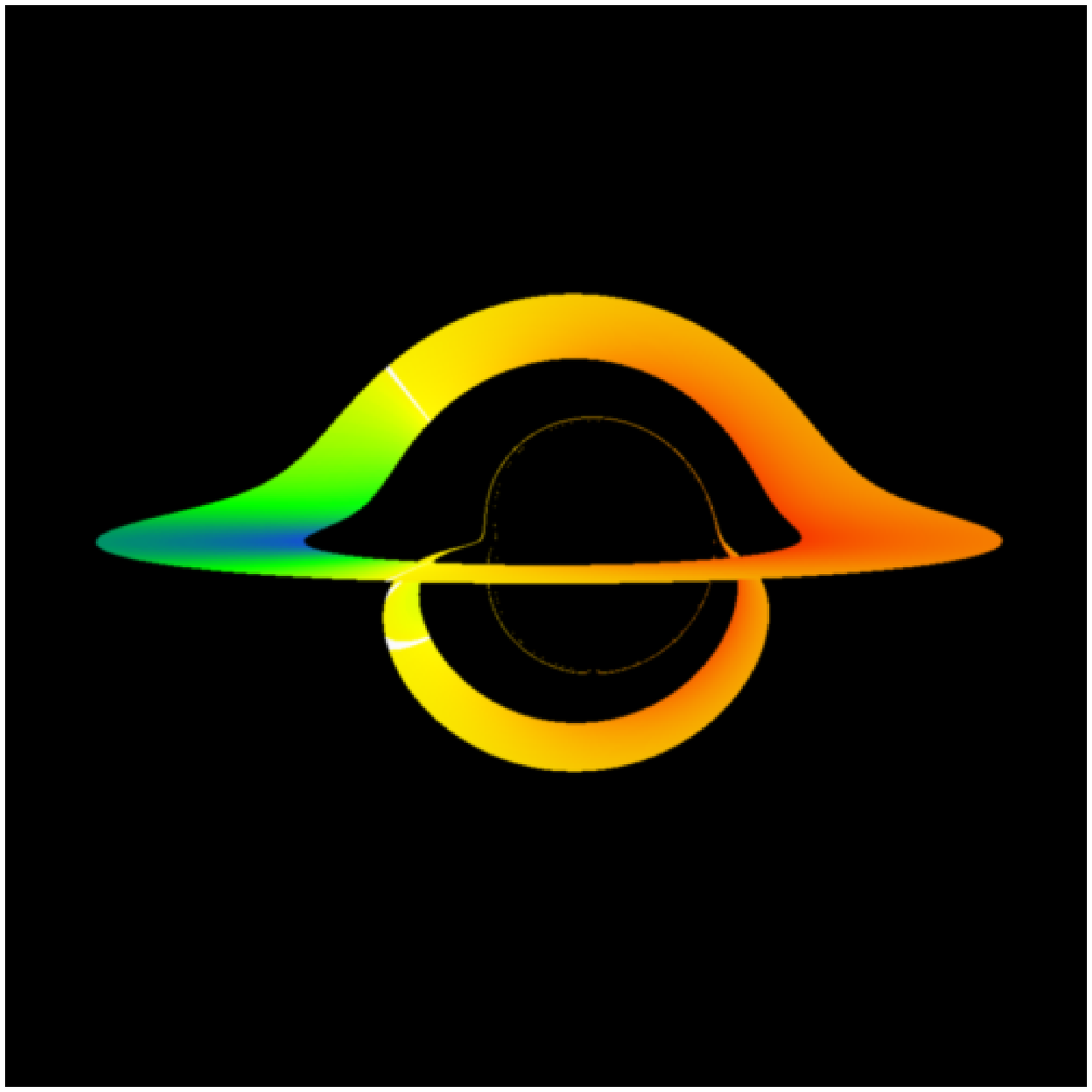,width=6.35cm}}
\center{\epsfig{figure=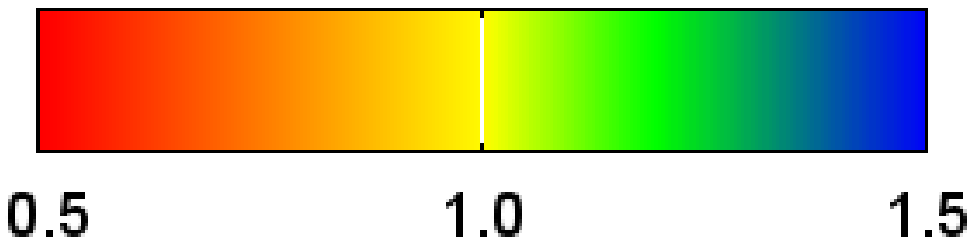,width=6.35cm}}
\caption{ 
     Energy shifts (with respect to a distant observer)   
        of geometrically thin accretion disks 
        with inner and outer boundaries the same as 
        the accretion torus with an aspect ratio  
        given by rotational velocity power-law indexs $n = 0.232$ 
        and Keplerian radius $r_{\rm k} = 12\RG$.   
    The spin parameter of the central black hole is  $a = 0.998$,   
    The viewing inclination angles are $i = 45^\circ$ (top panel) 
         and $85^\circ$ (bottom panel).  
   Blue represents energy blue shift and red represents energy red shift.
   The scale below the images shows the colour map used for relative energy
    shifts between 0.5 and 1.5
   The white shows the region where there is no energy shift. }
\label{image_disk}
\end{figure} 

\begin{figure} 
\vspace*{0.3cm}
\center{\epsfig{figure=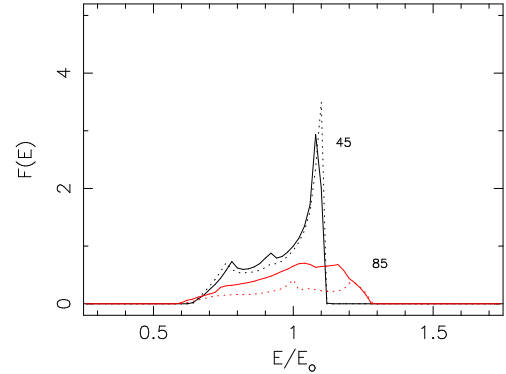,width=7.75cm}}
\caption{  Comparison of line profiles for  
    a geometrically thin accretion disk and an accretion torus 
    at viewing inclination angles $i=45^\circ$ and $85^\circ$. 
  The torus is that shown in middle and bottom panels of Fig.~\ref{image_4585rk}, 
      and the disk is that shown in Fig.~\ref{image_disk}.  
  The line profiles for the accretion torus are represented by solid lines 
     and the line profiles for the accretion disk are represented by dotted lines.  
  The line emissivity power-law index is $-2$ for both the torus and the disk. 
  The spin parameters of the central black hole is $a=0.998$. 
  The normalisation is such that $F(E) = 1$ at $E/E_{\rm o} = 1$ for the torus 
      viewed at $i = 45^\circ$. }
\label{v_dt}
\end{figure}

A thin accretion disk is practically a two-dimensional object, 
   in contrast to an accretion torus, which is three-dimensional. 
The whole upper surface of a flat accretion disk  is always visible 
   regardless of viewing inclinations, 
   but some fraction of the emitting surface of the torus will be self-eclipsed, 
  at sufficiently high view inclination angles.  
What aspects of the differences between accretion tori and accretion disks 
   give rises to different features in the profiles of their emission lines?     
Among the geometrical effects, which are the most important ones?   
Apart from geometrical factors, are there any different factors for accretion tori and disks  
  that cause differences in their line profiles?   
We now attempt to answer these questions.  
  
At moderate inclination angles, e.g. $45^\circ$, 
  if ignoring the small peak due to emission from high-order lensed images,   
  the torus lines are asymmetric and double-peaked, with profiles resembling those of the lines from relativistic thin disks.  
At large inclination angles, e.g. $85^\circ$, the torus lines are broad, asymmetric and single-peaked, 
   and so are the disk lines.  
Fuerst \& Wu (2004) showed a comparison of emission lines from a thin accretion disk 
   and emission lines from an accretion torus (given by rotational velocity power-law index $n = 0.21$).
See Figure ~\ref{old_dt} for a reproduction of the line profiles corresponding to this situation.  
Their central black holes have the same spin parameter $a=0.998$.     
The outer radii of the disk and the torus in the comparison are similar, roughly $20\RG$,
   and the line emissivity distributions of the torus and the disk all follow a radial power-law with an index of $-2$.  
Viewed at $i = 85^\circ$,  both the torus and disk lines are broad, asymmetric and single-peaked. 
The overall appearance of the disk line can be described as being wedge-shaped, 
   while the torus line is more like a hump.  
The most obvious difference between the two lines is the locations of the blue cut-off (edge/wing).    
The disk line has a sharp blue edge at 1.5 times the rest-frame line centroid energy, 
   while the torus line has a less steep blue wing, 
   with its flux falling off to zero at 1.25 times the rest-frame line centroid energy.
The red wings of the two lines are similar. 
However, the torus line does not extend as far as the disk line into the red.  

One may attribute the difference between the line profiles of the torus and the disk to the self-eclipse of the torus. 
If the most redshifted and the most blueshifted emissions from the inner torus regions are blocked,   
    the torus line will have weaker emissions at both red and blue line wings than the disk.   
An alternative explanation is the different kinematics for the emitters in the inner disk and inner torus regions.  
For the accretion disk considered in Fuerst \& Wu (2004), 
   the inner boundary is the last stable particle orbit set by the black hole spin. 
Its values is $1.23\RG$, corresponding to $a = 0.998$.  
The inner boundary surface of the torus is further out, located at $\approx 9.06\RG$ in the equatorial plane. 
Hence, the disk emissions would have larger redshift and larger blueshift than the torus emission.  
To disentangle these two factors, we need to compare lines from a torus and from a disk with the same inner and outer radii.
   
We choose the accretion torus with $n=0.232$ and $r_{\rm k} = 12\RG$ as the reference,    
   and construct an accretion disk with an inner radius and an outer radius the same as this accretion torus.  
Thus $r$ ranges from $9.0585 \RG$ to $17.7176\RG$. 
The projected images of the disk and the torus are very similar at moderate and small viewing inclination angles, 
   (cf.\   top panel of  Fig.~\ref{image_disk} and middle panel of Fig.~\ref{image_4585rk}), 
   but their projected images are very different  at high inclination angles 
   (cf. bottom panel  of Fig.~\ref{image_disk} and bottom panel of Fig.~\ref{image_4585rk}).    
Figure~\ref{v_dt} shows a comparison between the disk and torus line profiles. 
At $i=45^\circ$ the disk and torus lines roughly have the same profile.  
The disk line has a slightly stronger boosted blue peak and its red wing is slightly more extended. 
The central small peak of high-order lensed emission of the torus line 
   is slightly stronger and slightly redder than the disk line.      
The differences between the disk line and the torus line are more obvious  at $i = 85^\circ$. 
The disk line has less flux but more features than the torus line. 
The blue peak is present in the disk line but is not visible in the torus line. 
The central peak due to high-order lensed emission is also sharper and more visible in the disk line. 
The red and blue wings of the disk and torus line are almost identical. 
The disk line does not have a sharp edge, in contrast to the disk with an inner radius of $1.23\RG$.  
The blue wing shows a more gradual falling off, with flux reaching zero at roughly 1.28 of the rest-frame line centroid energy, 
   which is almost identical to that of the torus line.  

This demonstrates that 
   if the disk and the torus have similar inner and outer radii, 
   the lines from their surface can barely be distinguished at moderate or small viewing inclination angles. 
At very large inclination angles, 
   the differences between the disk and torus line manifest mostly in the total line flux 
   and in weakly energy-shifted central regions of the line profiles. 
For accretion tori, the line wing morphology is not sensitive to whether  self-eclipse occurs or not. 
Its dependence to projection (due to viewing inclination) is similar to that of the disk lines. 
We  may therefore conclude that 
   the great differences in the wings of the torus line and the disk line seen in Fuerst \& Wu (2004) 
   is caused by different kinematics for the emitters in their innermost regions, 
   as the torus and the disk have different extensions of the inner boundary toward the central black hole.   
For the same reason, the weak dependence of the profiles of torus lines on the black hole spin  
   shown in Figure~\ref{v_spin} can be explained by 
   the difference in the inner radii of the tori being insignificant. 
      
These conclusions hold if we assume that the emissions are from the surfaces 
  of the disks and the tori. 
If the torus and the disk are transparent to the emissions, 
   the emissions are also weighted by the interior structures of the emitting object. 
A transparent (or semi-opaque) torus, which is  a three-dimensional object, and a transparent (or semi-opaque) disk    
   would show differences in their lines because of other effects such as optical depth and differential kinematics
   (Fuerst 2005).    
   
\subsection{Emission from accretion tori with more extreme parameters}    

In previous sections we considered accretion tori that have a velocity law similar to that 
   obtained by numerical simulations, 
   but we have set the Keplerian radius to large values ($r_{\rm k} = 8$ and $12\RG$).   
This, together with the prescription that we have adopted for an inner boundary emitting surface of the tori, 
  implies that emissions are from regions relatively far from the black-hole event horizon (see Fig.\ \ref{image_4585rk}).   
One may expect that the resulting emissions would suffer smaller gravitational red-shift 
   than the emissions for disk/torus models with the innermost emission surface closer to the black-hole event horizon. 
The remaining question is now: Is this argument valid where the accretion torus has substantial thickness close to the black hole?
  
\begin{figure} 
\vspace*{0.5cm}
\center{\epsfig{figure=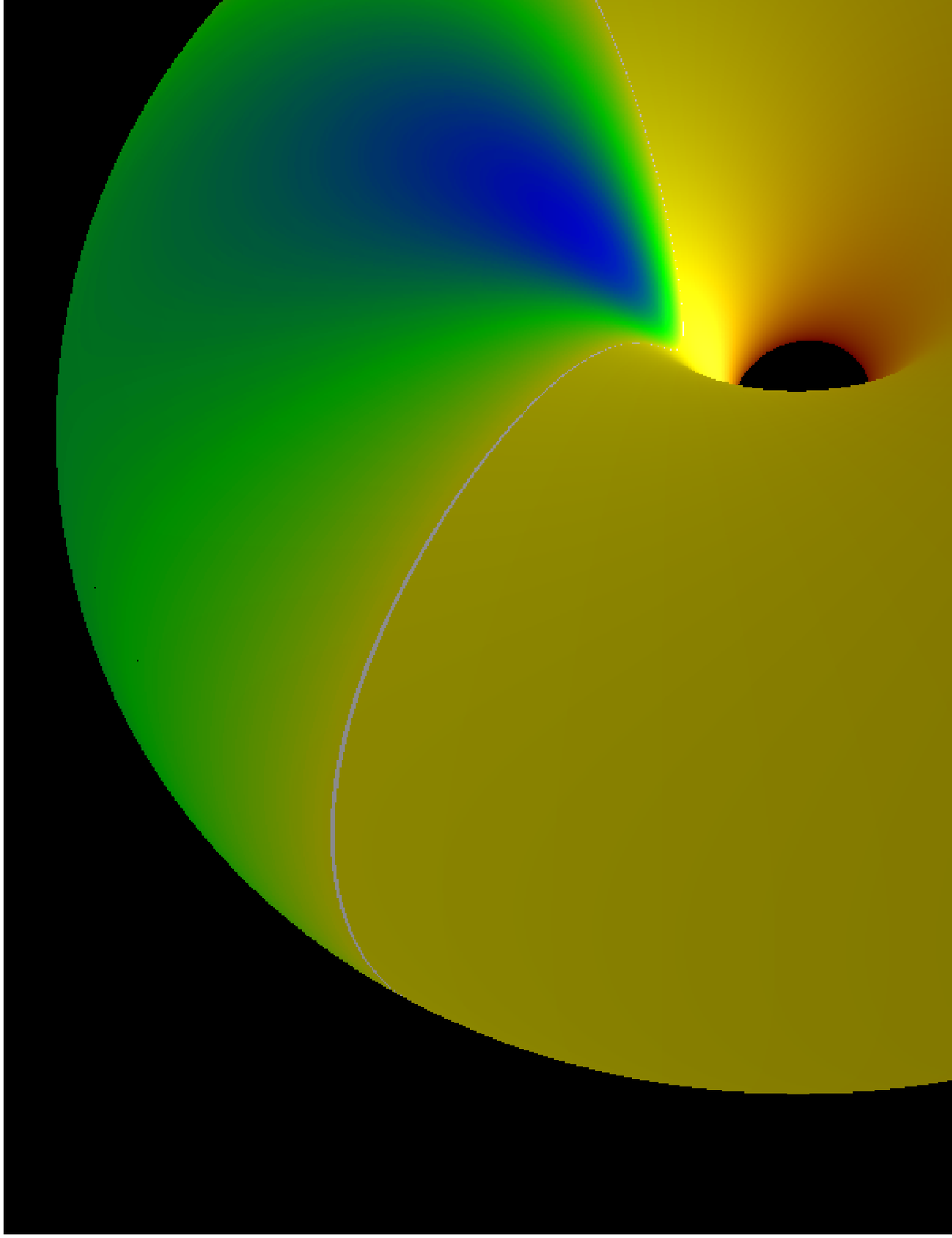,width=6.35cm}}
\center{\epsfig{figure=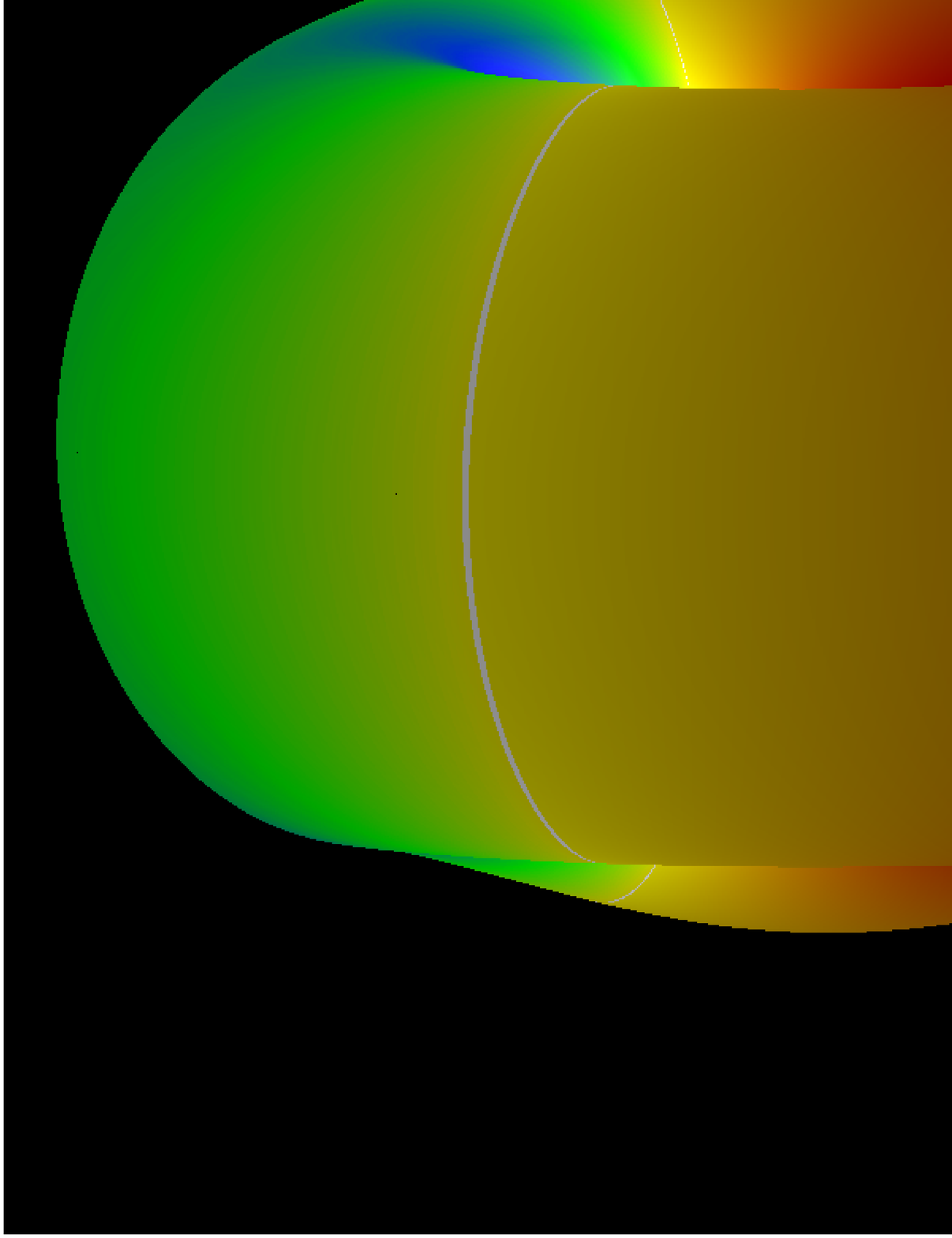,width=6.35cm}}
\center{\epsfig{figure=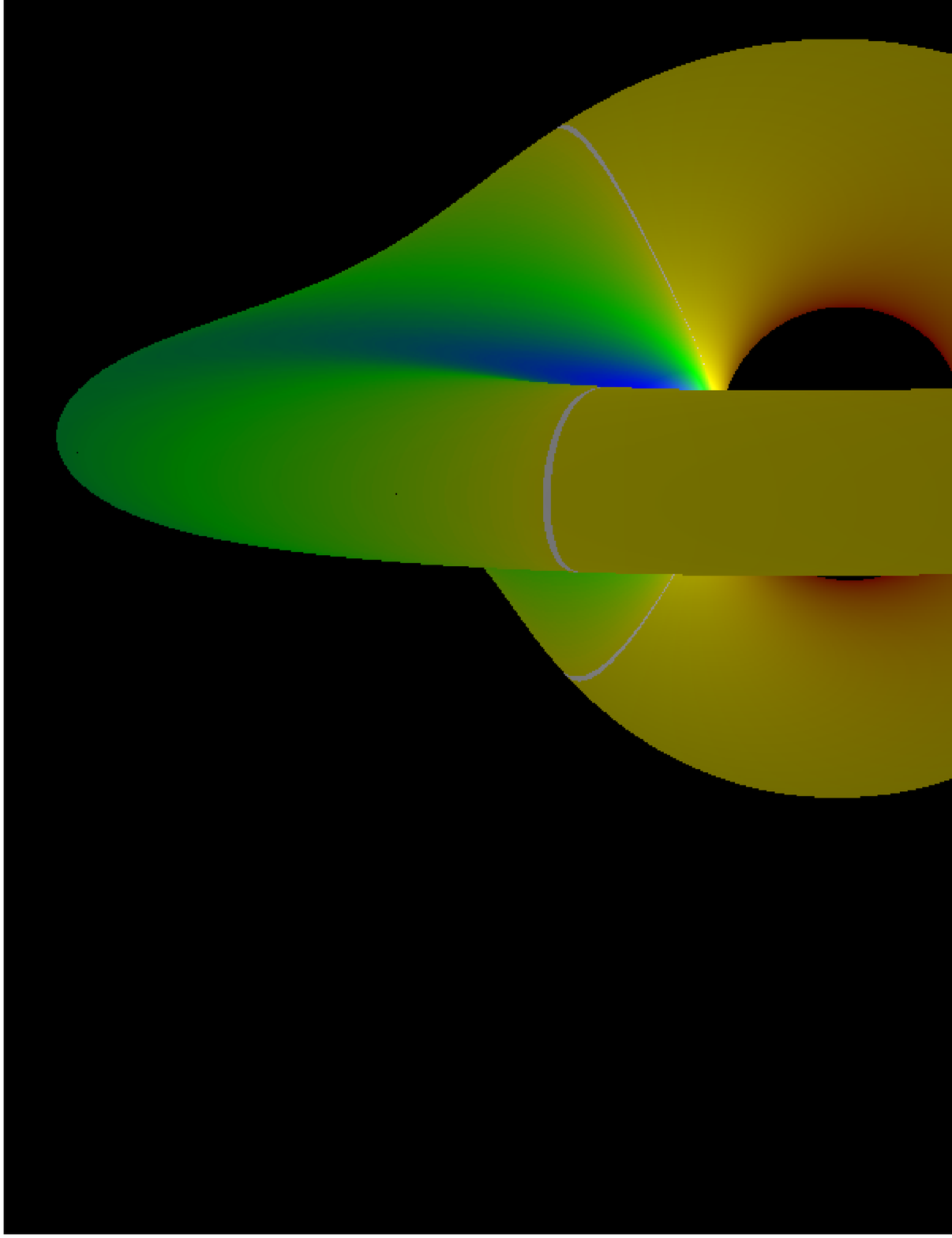,width=6.35cm}}
\caption{ 
The energy shifts (with respect to a distant observer) of the emission 
   from accretion tori with the inner boundary of the emission surface 
   reaching the black-hole event horizon.  
The Keplerian radius of the tori is $r_{\rm k} = 1.3 \RG$,  
   and the emission surface is assumed to be an isobaric surface 
   crossing $20\RG$ on the equatorial plane.  
The black-hole spin parameter is $a=0.998$.  
The viewing inclination angles of the tori of the top and middle panels  
   are $i=45 ^\circ$ and 85$^\circ$ respectively.  
The velocity law is given by $n=0.2$. 
For the torus in the bottom panel, 
  $i=85^\circ$  and $n=0.01$. 
 }
\label{image_flat_tori}
\end{figure} 

\begin{figure}
\vspace*{0.3cm} 
\center{\epsfig{figure=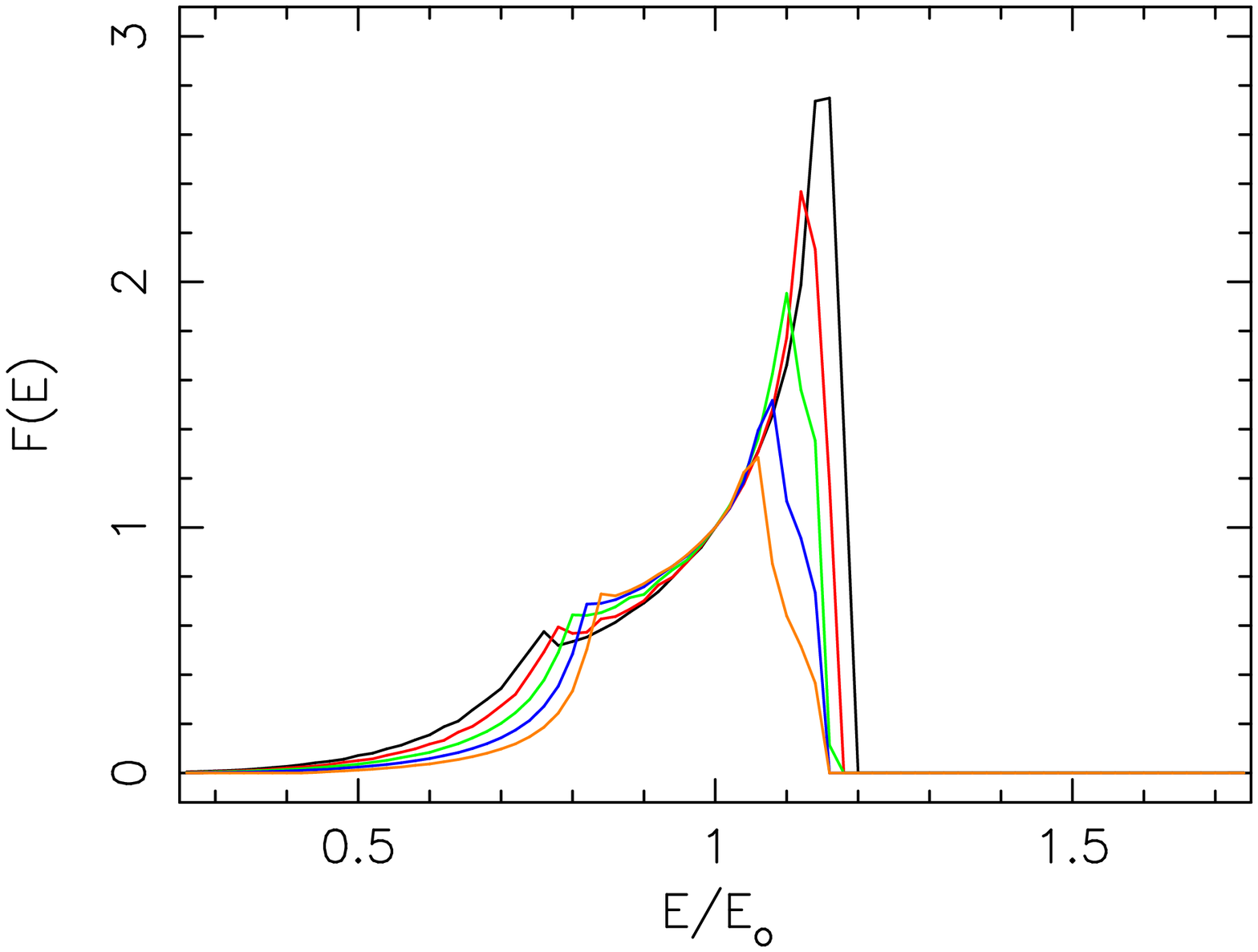,width=7.75cm}} 
\center{
\tiny
\begin{tabular}{|c|ccccc|}
\hline
$n$&0.01&0.05&0.1&0.15&0.2\\
$r_{\rm in}$&1.3050&1.2855&1.2824&1.2752&1.2666\\
$z_{\rm max}$&2.5145&5.5872&7.8263&9.4780&10.805\\
aspect ratio&3.7174&1.6748&1.1958&0.98780&0.86688\\
\hline
\end{tabular}}
\caption{
Table of parameters used for the torus model,   
  with $r_{\rm max}=20\RG$, $a=0.998$ and $r_{\rm k} = 1.3\RG$. 
The aspect ratios of the tori are determined by  $n$,  
   which takes the values of 0.01, 0.02, 0.10, 0.15 and 0.20,
where the aspect ratio is defined by the partial width ($r_{\rm max} - r_{\rm in}$) divided by the total height ($z_{\rm max}\times2$) of the tori.
The profiles of the emission lines are shown in the plot, 
  assuming an emissivity power law index of $-2$ and a viewing inclination of $60^\circ$. 
The line with the largest width correspond to the torus with $n=0.01$, 
  and the line width decreases when $n$ increases.    
The lines are normalized such that  $F(E) = 1$ at $E/E_{\rm o} = 1$.  }
\label{torus_aspect_table}
\end{figure}

The emission surface is not necessarily coincident with the critical surface 
   that passes through the last stable orbit on the equatorial plane.
In other words, the surface at which the line emission has unit optical depth 
   may be significantly closer to the black-hole event horizon than the outermost isobaric surface. 
Here we relax the assumption that the emission surface is the isobaric surface intercepting the marginally stable orbit.  
We consider an illustrative case which sets $r_{\rm k} = 1.3 \RG$.  
The central black hole has a spin-parameter $a=0.998$. 
We assume that the isobaric surface crossing the equatorial plane at $20\RG$ is the line emission surface. 
This gives an inner boundary for the emission surface almost reaching the black-hole event horizon.   
We alter the aspect ratio of the torus by varying its rotation law power-law index $n$.  
This process does not significantly alter the radius of the innermost emission boundary surface of the torus.
It therefore allows us to disentangle the effects due to the torus geometry on the line profiles  
   from those due to gravity and relativistic motion. 
Figure~\ref{torus_aspect_table} shows the lines from tori with different aspect ratios,  specified by $n$, 
  and viewed at $i = 60^\circ$. 
The line width clearly changes with $n$, and at this viewing angle, 
   the thicker the tori (larger $n$) the narrower the line. 
Moreover, the blue line peak is suppressed when $n$ increases.   
   
\begin{figure} 
\vspace*{0.3cm}
\center{\epsfig{figure=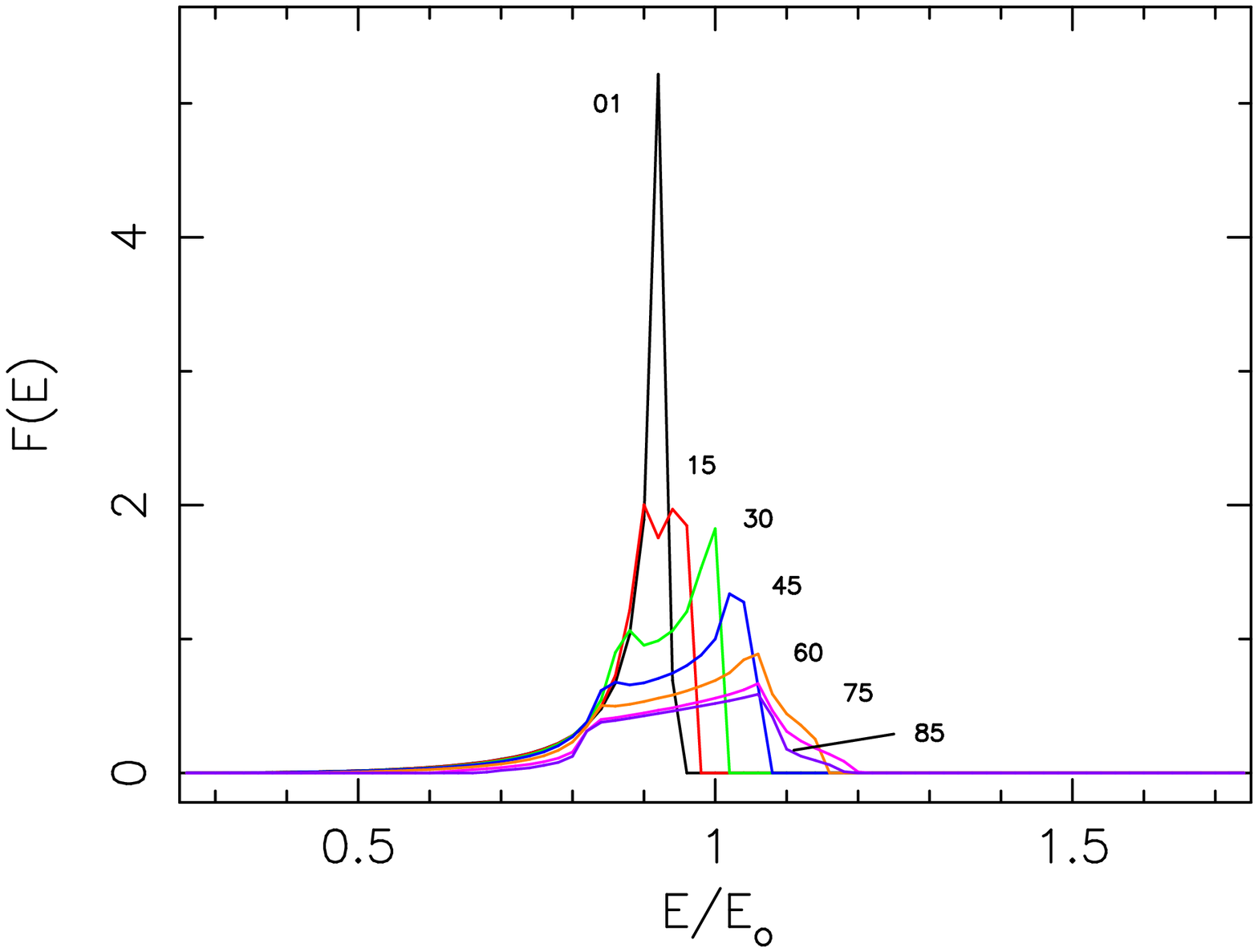,width=7.75cm}} 
\center{\epsfig{figure=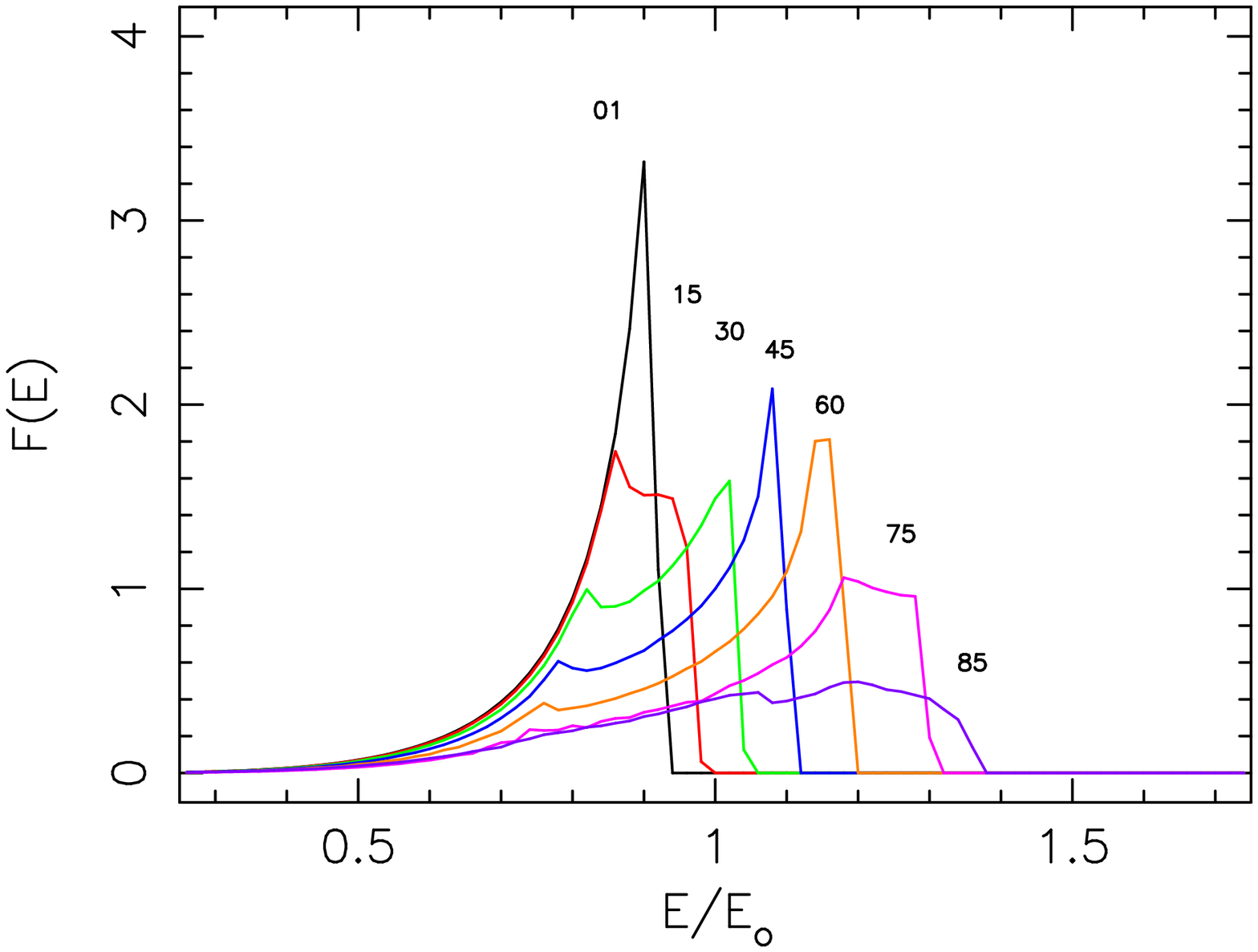,width=7.75cm}}
\caption{
  Profiles of emission lines from accretion tori viewed at inclination angles 
      $i = 1^\circ$,  $15^\circ$, $30^\circ$,  $45^\circ$, $60^\circ$, $75^\circ$ and  $85^\circ$.  
  The radial emissivity power-law index is $-2$.  
  The radius of Keplerian rotation is $r_{\rm k} = 1.3\RG$, 
     and the spin parameter of the central black holes is $a= 0.998$.  
  The rotational velocity power law indices are $n=0.2$ (top) and 0.01 (bottom).  
  The normalisation is such that $F(E) = 1$ at $E/E_{\rm o} = 1$ for $i=45^\circ$.   }
\label{line_flat_tori}
\end{figure}

\begin{figure} 
\vspace*{0.3cm}
\center{\epsfig{figure=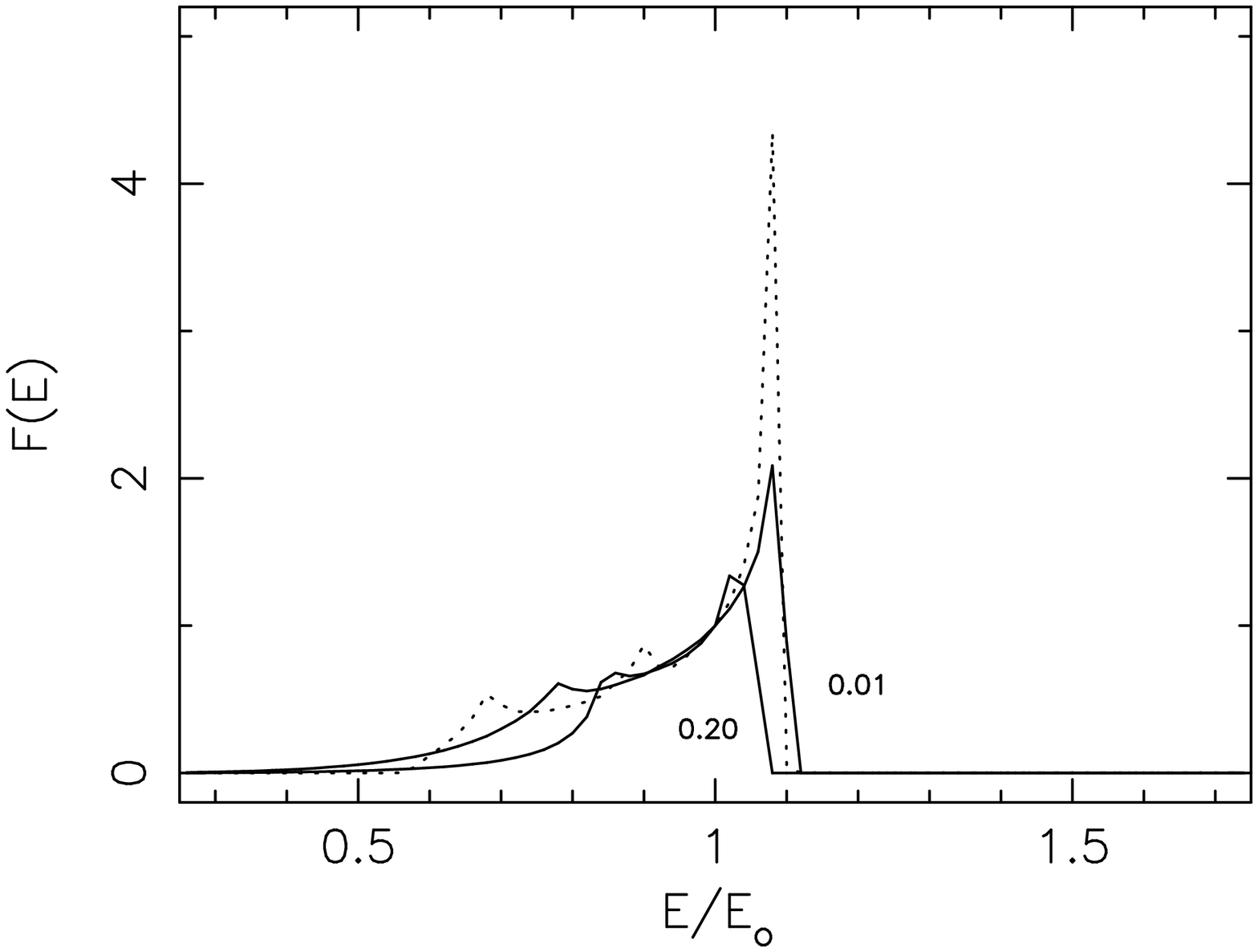,width=7.75cm}} 
\center{\epsfig{figure=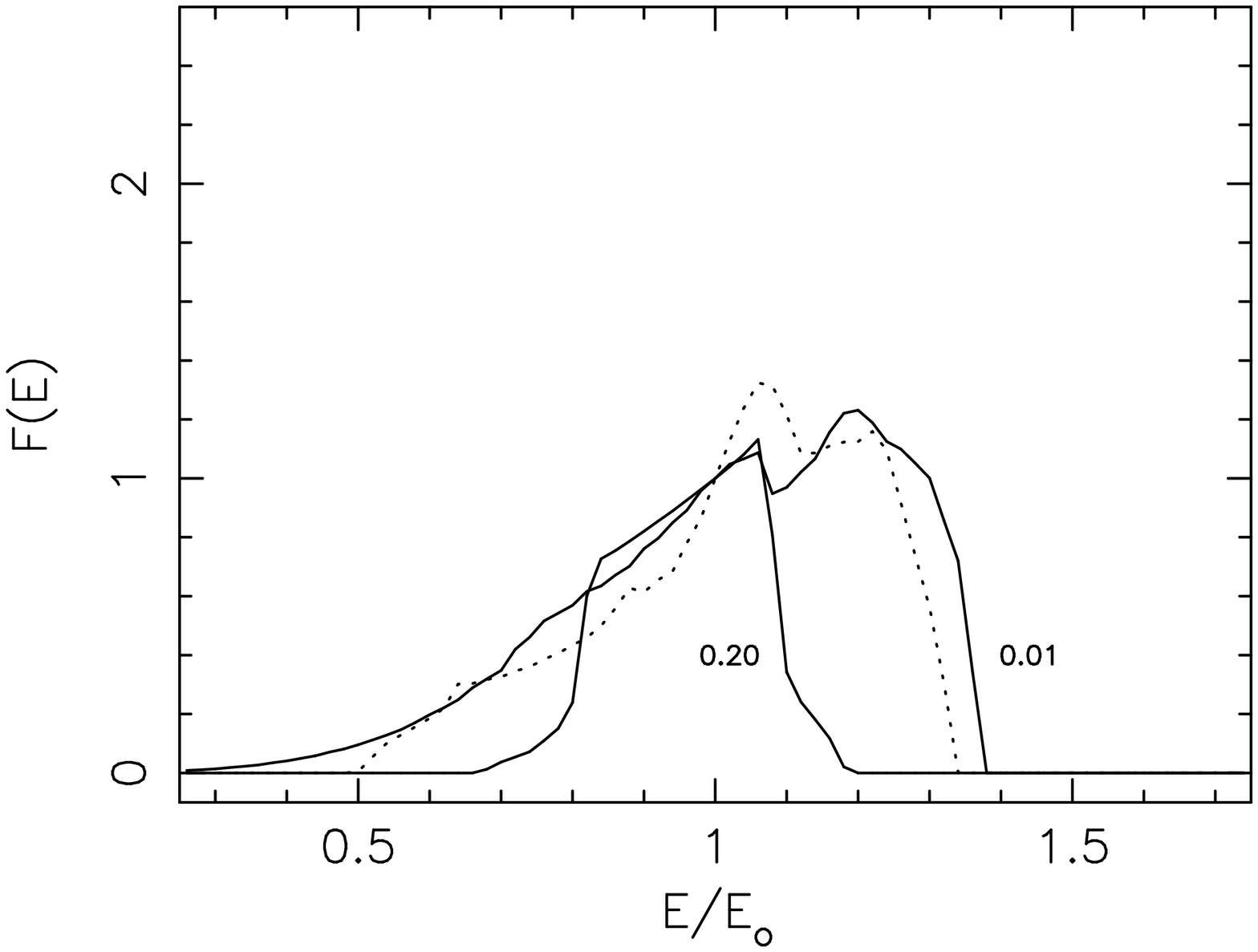,width=7.75cm}}
\caption{
  Profiles of emission lines from accretion tori around black holes with $0.998$ 
      viewed at an inclination angle $i=45^\circ$ (top) and $85^\circ$ (bottom). 
  The radial emissivity power-law index is $-2$.  
  The radii of Keplerian rotation are $r_{\rm k} = 1.3\RG$ (solid lines) and 8$\RG$ (dotted line).  
  For the tori with $r_{\rm k} = 1.3\RG$, 
     the rotational velocity power law indices are $n=0.2$ and 0.01 
     (as marked respectively); 
     for the torus with $r_{\rm k} = 8\RG$, $n=0.232$.  
  The normalisation is such that $F(E) = 1$ at $E/E_{\rm o} = 1$.     }
\label{line_tori_compare}
\end{figure}

In what follows, we compare the thickest ($n=0.2$) and thinnest ($n=0.01$) models.
Figure~\ref{image_flat_tori} shows examples of such accretion tori. 
There are several noticeable differences between these extreme accretion tori 
   and those discussed in the earlier section. 
Firstly, the inward extension of the unit optical-depth surface to the black-hole event horizon 
   will block all high-order emission passing through the inner hole of the torus.  
There will be no high-order emission unless the torus is viewed at very high inclinations, 
   such as $i >80^{\circ}$ depending on the torus aspect ratio.   
Secondly, self-occultation is more severe in the tori considered here, 
   and the degree of occultation increases with the viewing inclination 
   (cf.\  top and middle panels, Fig.~\ref{image_flat_tori}) 
  and with thickness of the torus (cf.\  middle and bottom panels, Fig.~\ref{image_flat_tori}).  
For the flat geometrically thin accretion disks, the upper disk surface is always visible to a distant observer. 
Emission from the upper disk surface and from the inflowing disk-fed material inside the last-stable particle orbit 
   are always visible to the distant observer. 
For the accretion tori discussed in the previous sections, 
   some part of the emission region (but not all) in the upper plane is obscured by the torus 
   because it has thickness.  
However, for the extreme tori that we constructed here 
   the innermost emission region can be completely invisible to the observer. 
The emission from the gas closest to the black hole will have the largest gravitational red-shift 
   and it will also have the largest Doppler shift and boost.   
These occultation effects are seen in the profiles of the line emission from the tori.    
   
Figure~\ref{line_flat_tori} shows the line profiles for the two extreme tori with parameters 
   like those in Figure~\ref{image_flat_tori} 
   viewed at different inclinations. 
In comparison with the lines in Figure~\ref{v_spin},  
   the lines from the two tori generally have larger red-shift at small inclination angles. 
This is expected, as we allow the emission to arise from regions very close to the black hole event horizon, 
   and at these inclinations the emission is not blocked. 
The lines from the (thinner) more disk-like torus with $n=0.01$ generally have both larger red-shift and larger blue-shift 
   than the (thicker) torus with $n=0.2$, 
   which is simply because the inner region of the former torus is more visible than that of the latter torus. 
As an illustration of the complex interplay between viewing and geometrical aspect, 
   we show the comparison of the normalised profiles of the lines from these two tori to the the lines 
   from the torus with $r_{\rm k} = 8\RG$ and $n=0.232$ 
   (Fig.~\ref{line_tori_compare}). 
At moderate viewing inclinations, say $i \approx 45^\circ$ (top panel, Fig.~\ref{line_tori_compare}),  
   the line profiles of the two extreme tori appear to be narrower than the torus with large $r_{\rm k}$, 
   in spite of the inner emission surface of these tori being much closer to the black-hole event horizon.  
Also, the blue peak of the line is less boosted, because the most blue-shifted emission is obscured. 
This demonstrates that geometrical effects can be important when the accretion torus/disk has non-negligible thickness. 
At high viewing inclination, say $i> 85^\circ$, the situation becomes ambiguous. 
If the torus is very thick, the line can be comparably narrow.     
As the innermost region is no longer visible to the observer,  
  the emission is contributed mostly by the outer torus surface, 
  where the strong relativistic effects are 
  line broadening caused by Doppler motion of the emitters and red-shift due to the transverse Doppler effect (time dilation). 
Thus, the dominant effects are (kinematic) special relativistic instead of (gravitational) general relativistic. 
For geometrically thinner tori (e.g. with $n=0.01$) the results are more disk-like, and 
  general relativistic effects can play some role, 
  as shown in Figure~\ref{line_tori_compare} (bottom panel).
        
\subsection{Astrophysical implications} 

Several AGN (e.g.\ MCG-6-30-15) have been found 
   to show broad, asymmetric and double-peaked Fe K$\alpha$ lines in their keV X-ray spectra. 
There were occasions when the line had a profile closely resembling those of relativistic disks. 
It has been suggested that we can use the line profiles to determine various system parameters 
   such as the black hole spin.  
However, is puzzling that observations of the same sources at another epochs 
   could show different profiles for the same Fe K$\alpha$ lines. 
Calculations have shown that geometrically thin relativistic disks have characteristic asymmetric line profiles.  
If internal and external absorption are unimportant, 
  and if the lines are not contaminated by a underlying continuum with edge features, 
  the general shape of the line profiles is quite robust.
  
Our calculations here show that the torus lines have certain properties similar to the disk lines.  
Moreover, our calculations further show that unless the viewing inclination angle is very large,  
   if the inner boundary is far from the black hole, 
   the profiles are not very sensitive to other system parameters.
However, if the inner edge of the torus is close to the black hole, the line wings sensitively depend on the accretion torus aspect ratio.
Thus, if an AGN shows variations in the line profiles at different observational epochs, the overall shapes of the line may allow one to deduce the changes in the accretion flow.

These results also show that disentangling the effects of the black hole spin on line profiles is complex when the geometry is no longer a thin disk whose inner radius is delineated by the marginal stable orbit.
Line profiles from tori may emulate those from disks with vastly different inclination, black-hole spin, and inner radius.

\section{Conclusion}

We constructed model accretion tori 
   and calculated the profiles of emission lines from them, 
   assuming that the tori are optically thick to the lines,  
   so the lines are emitted from a thin surface.  
We have shown the the line profiles vary with viewing inclination, 
    from narrow single-peaked lines at low inclinations, to asymmetric double-peaked lines at moderate inclinations 
    to broad asymmetric single-peaked lines at high inclinations.  
The line profiles also depend on the location of the inner and outer boundaries of the torus. 
They are not sensitively dependent on the spin of the central black hole, 
    as the inner boundary of  the torus is set by the balance of the forces due to pressure and gravity 
    instead of by the last stable Keplerian orbit.   
Self-eclipse and lensing play some role in shaping the torus line at high aspect ratios.
For accretion tori, we may use the lines to constrain the viewing inclination of the system 
   and the inner and outer boundaries of the torus. 
Using the line profiles to constrain other parameters is less reliable.

\begin{acknowledgements}
We thank John Peterson for carefully reading the manuscript and providing suggestions.   
We also thank the referee for critical comments and helpful suggestions.       
 
\end{acknowledgements}

\end{document}